\documentclass[a4paper,fleqn,usenatbib]{mnras}

\usepackage[T1]{fontenc}
\usepackage{ae,aecompl}
\usepackage{relsize} 
\usepackage{color,soul}

\usepackage{graphicx}	
\usepackage{amsmath}	
\usepackage{amssymb}	
\usepackage{bm}
\usepackage{footnote}

\title[PPD Truncation Mechanisms]{Protoplanetary disc truncation mechanisms in stellar clusters: comparing external photoevaporation and tidal encounters}

\author[Winter et al.]{A.~J.~Winter$^{1}$,\thanks{E-mail: ajwinter@ast.cam.ac.uk} C.~J.~Clarke$^{1}$, G.~Rosotti$^{1}$, J.~Ih$^{2}$, S.~Facchini$^{3}$, T.~J.~Haworth$^{4}$
\\
$^{1}$Institute of Astronomy, Madingley Road, Cambridge CB3 0HA, UK \\
$^{2}$Clare College, Trinity Lane, Cambridge, CB2 1TL\\
$^{3}$Max-Planck-Institut f{\"u}r Extraterrestrische Physik, Giessenbachstrasse 1, D-85748 Garching, Germany \\
$^{4}$Astrophysics Group, Blackett Laboratory, Imperial College London, Prince Consort Road, London SW7 2AZ, UK
}

\date{Accepted XXX. Received YYY; in original form ZZZ}

\pubyear{2017}

\begin{document}
\label{firstpage}
\pagerange{\pageref{firstpage}--\pageref{lastpage}}
\maketitle

\begin{abstract}
Most stars form and spend their early life in regions of enhanced stellar density. Therefore the evolution of protoplanetary discs (PPDs) hosted by such stars are subject to the influence of other members of the cluster. Physically, PPDs might be truncated either by photoevaporation due to ultraviolet flux from massive stars, or tidal truncation due to close stellar encounters. Here we aim to compare the two effects in real cluster environments. In this vein we first review the properties of well studied stellar clusters with a focus on stellar number density, which largely dictates the degree of tidal truncation, and far ultraviolet (FUV) flux, which is indicative of the rate of external photoevaporation. We then review the theoretical PPD truncation radius due to an arbitrary encounter, additionally taking into account the role of eccentric encounters that play a role in hot clusters with a 1D velocity dispersion $\sigma_v \gtrsim 2$~km/s. Our treatment is then applied statistically to varying local environments to establish a canonical threshold for the local stellar density ($n_\mathrm{c} \gtrsim 10^4$~pc$^{-3}$) for which encounters can play a significant role in shaping the distribution of PPD radii over a timescale $\sim 3$~Myr. By combining theoretical mass loss rates due to FUV flux with viscous spreading in a PPD we establish a similar threshold for which a massive disc is completely destroyed by external photoevaporation. Comparing these thresholds in local clusters we find that if either mechanism has a significant impact on the PPD population then photoevaporation is always the dominating influence. 
\end{abstract}

\begin{keywords}
accretion, accretion discs -- protoplanetary discs -- circumstellar matter -- stars:
kinematics and dynamics -- stars: pre-main-sequence
\end{keywords}



\section{Introduction}

The majority of stars form from dense cores in giant molecular clouds (GMCs), and therefore young stars tend to spend the early phases of their evolution in regions of enhanced stellar density \citep[e.g.][]{Lada03}. If this environment is sufficiently dense, a protoplanetary disc (PPD) can undergo truncation and mass loss due to close encounters \citep{Cla93, Ost94,Hal96,Pfa05,Olc06, Bre14,Mun15, Win18} and external photoevaporation \citep{Sto99, Arm00, Sca01, Ada10,Fac16, Gua16}. The physical importance of these processes for disc evolution remains an open question, and is likely to rely on the early stages of cluster evolution. For example, if stars form with a subvirial velocity dispersion, as suggested by observations \citep{Tob09}, the cluster will undergo cold collapse, enhancing encounter rates. However, if star formation efficiency (SFE) is low in a molecular cloud, as is expected from observations \citep{Lada03} and simulations \citep[e.g.][]{Mur11, Pfa13}, the cluster can become supervirial subsequent to gas expulsion, leading to dispersal of stars into the field \citep[though see][]{Kru12b}. The presence of sub-structure in the cluster \citep[e.g.][]{Wil94, Bon03} can - despite the fact that it is lost over a crossing time - also induce higher degrees of disc truncation \citep{Par11, Cra13}. 

Our focus here is on the distribution of outer disc radii within a cluster, a physical property that has only recently become measurable. With ALMA, observations of PPDs which are sufficiently spatially resolved ($\sim 0.2-0.3''$) to estimate the outer radii at (sub-)mm wavelengths are becoming available, and samples sizes of $\sim 10-100$ discs have been collated \citep{Ans16,Ans17, Ans18, Bar17, Cox17, Taz17, Tri17}. However, radial extent estimates are almost exclusively based on the (sub-)mm continuum brightness distribution, which traces the dust content of the disc. In the context of measuring truncation effects, this is problematic as observations suggest that gas extends to larger radii than the dust, either due to gas lines being more optically thick than the (sub-)mm continuum \citep{Dut98, Gui98, deG13} or the radial drift of dust particles \citep{And12, Pie14}. Spatially resolved measurements of the gas are more challenging due to reduced flux and spatial resolution, a problem which is further compounded by the dependence of gas temperature, and hence radial intensity profiles, on the properties of the dust \citep{Fac17}. However, as significant progress has been made in the last few years towards this goal, it may be possible to test hypotheses on disc radii distributions in the near future.

In this paper we perform a comparative study of the roles of tidal truncation and photoevaporation in setting the  distribution of protoplanetary disc radii in clustered environments. To this end we compile a census of well studied star forming regions in Section \ref{sec:clust_environs} and depict them in the plane of ultraviolet field strength versus stellar density; these being, respectively, the main parameters that determine the importance of photoevaporative effects and dynamical truncation. This work bears closest similarity to that of \citet{Ada06}, who focused on the early stages of dynamical cluster evolution on young planetary systems. In that work both close encounters and FUV flux were considered. However, the intention was more on quantifying the dynamical evolution of young clusters, and less on estimating the resulting disc properties. In this work we will instead take observed environments as `snapshots' in which we quantify the influence of PPD truncation mechanisms. Additionally \citet{Ada06} only considered small clusters of $100-1000$ stars. Practically this means that these clusters are unlikely to host massive stars, and therefore FUV flux throughout the environment is low. In reality the mass function for young clusters is not steep below $\sim 10^5- 10^6 \, M_\odot$ \citep{Sch76, Gie06,Bas08, Zwa10}, and we therefore expect a large fraction of stars to spend their early phases in much more populated environments. This work addresses the properties of observed clusters in the higher mass limit, and additionally contributes to the quantification of the truncating effects of photoevaporation and tidal encounters.

To establish the role of cluster density and local FUV flux we present some further theoretical development of the two truncation scenarios.  We explain in Section \ref{sec:tidal} why, despite the large body of previous work on dynamical disc truncation in the literature, both for individual encounters \citep[e.g.][]{Hal96, Pfa05, Bre14, Bha16} and in terms of clustered stellar populations \citep{Sca01, Olc06,Pfa06,Ada06, Olc10, Cra13, Vin16}, it is necessary to perform some further simulations to explore regimes involving hyperbolic orbits and a large dynamic range of stellar masses. We use expressions fitted to these numerical results to assess the average truncation radius of stars that remain in an environment of fixed stellar density over a given time interval. Likewise in Section \ref{sec:photoevap} we perform similar calculations in the case of photoevaporation. These calculations differ form previous works in that they consider a wider range of ultraviolet field fluxes \citep[cf.][]{Cla07,And13} and take into account the viscous evolution of the disc \citep[cf.][]{Joh98, Ada04}; the calculations presented here bear closest resemblance to those of the study of photoevaporation of discs in very low mass stars by \citet{Haw18}. Here however our focus is on the $\sim 1\, M_\odot$ stellar regime and in particular we focus on disc radius distributions in order to compare with the results of dynamical truncation presented in Section \ref{sec:tidal}. Comparisons between the two truncation mechanisms are drawn in Section \ref{sec:comparison}. Concluding remarks are made in Section \ref{sec:conclusion}.

\section{Cluster Environments}
\label{sec:clust_environs}

Our first stage in producing comparisons between truncation mechanisms is to assess the local conditions within observed real clusters, which we assume are representative of stellar populations. We aim to produce a distribution of the far ultraviolet (FUV) flux in terms of the interstellar value $G_0 \equiv 1.6 \times 10^{-3}$~erg~s$^{-1}$~cm$^{-2}$ for real stellar cluster members \citep{Fat08}, and the corresponding local stellar number densities such that an estimation of the outer radius evolution can be made. To that end we discuss the FUV luminosity as a function of stellar mass, and modelling assumptions for real clusters.

\subsection{Properties of Stellar Clusters}
\label{sec:clustMC}

\begin{figure}
	\includegraphics[width=\columnwidth]{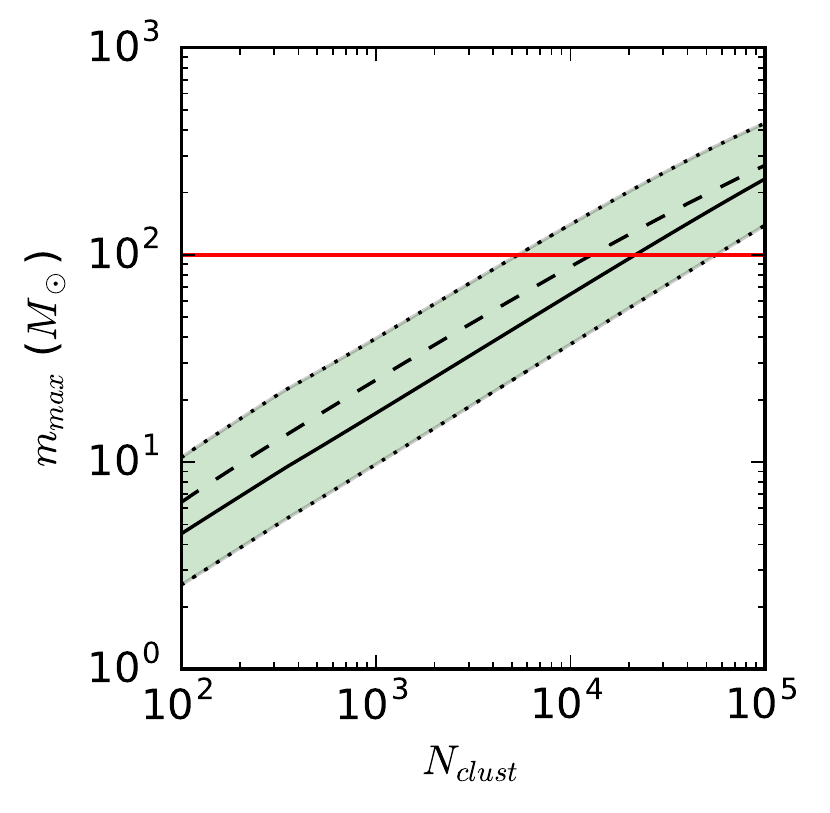}
    \caption{The mass of the most massive cluster member $m_\mathrm{max}$ as a function of the number of members of that cluster $N_\mathrm{clust}$. The solid line is the median $m_{1/2}$ and the dashed line is the mean $\bar{m}_\mathrm{max}$. The dotted line represent the $1$-$\sigma$ range, which is shaded. The horizontal red line at $100\, M_\odot$ is the greatest mass for which our stellar atmosphere models apply, and therefore an effective upper limit on the $m_\mathrm{max}$.}
    \label{fig:pmmax}
\end{figure}

\begin{table*}

    \begin{minipage}{\linewidth}
        \renewcommand\footnoterule{}
        \renewcommand{\thefootnote}{\alph{footnote}}
\begin{center}
 \begin{tabular}{c | c c c c c c c c} 
 \hline
Cluster  &  $\rho_0$ ($M_\odot$~pc$^{-3}$) & $r_\mathrm{core}$ (pc) & $r_\mathrm{eff}$ (pc) & $r_\mathrm{t}$ &  $\gamma$ & $M_\mathrm{clust}$ ($M_\odot$)  & $m_{\mathrm{max,}}^\mathrm{-\sigma}$ ($M_\odot$) & $m_\mathrm{max}^\mathrm{obs}$ ($M_\odot$) \\ [0.5ex] 
 \hline\hline
 NGC 3603 & $1.05 \cdot 10^5$ & $0.15$ & $0.7$ & $3.41$ & $2.00$ & $1.3 \cdot 10^4$ & $67$ & -  \\
Trumpler 14 & $1.25 \cdot 10^5$ & $0.14$ & $0.5$ & $1.92$ & $2.00$ & $10^4$ & $58$ \\
ONC& $1.03 \cdot 10^4$ & $0.2$ & $2.0$ & $20.18$ & $2.00$ & $4.5 \cdot 10^3$ & $37$ & $\sim 37$ \\
Arches& $1.30 \cdot 10^5$ & $0.2$ & $0.4$ & - & $3.27$ & $2.00 \cdot 10^4$ & $87$ & - \\
Quintuplet& $523$ & $1.0$ & $2.0$ & - & $3.27$ & $10^4$ & $58$ & - \\
Wd 1\footnotemark[1]  & $9.52 \cdot 10^4$ & $0.28$ & $0.86$& - & $4.00$ & $3.2 \cdot 10^4$ & $114$  & -
\\ \hline
Cygnus OB2 & $21.9$ & $3.9$ & $5.1$ & - & $5.80$ & $1.7 \cdot 10^4$ & $78$ & $\sim 100$ \\
Serpens A & $743$ & $0.16$ & - & $0.25$ & $4.00$  & $17$  & - &$5.1$ \\
Serpens B & $495$ & $0.14$ & - & $0.21$ & $4.00$  & $6.8$  & - &$5.1$\footnotemark[2] \\
$\sigma$ Ori & $542$ & $0.17$ & $0.41$ & $3.00$ & $1.30$ &  $146$ & $5.1$ & $17$ \\ 
$\lambda$ Ori& $106$ & $0.33$ & $2.96$ & $14.00$ & $1.80$ & $214$ & $6.4$ & $26.8$ \\
NGC 2024 & $2.16 \cdot 10^3$ & $0.16$ & $0.24$ & $0.90$ & $4.01$ & $132$ & $4.8$ & $15-25$ 
\\ [1ex] 
 \hline
\end{tabular}
\vspace{-1.5ex}

\footnotetext[1]{ Although numbers are recorded in \citet{Zwa10}, the reported values for $r_\mathrm{core}$, $r_\mathrm{eff}$ and $\gamma$ are inconsistent. We therefore use $r_\mathrm{eff}$ from \citet{Men07}, and fit an appropriate core radius.}
\footnotetext[2]{ This is the maximum mass found throughout Serpens, placed at the centre of Serpens A. Therefore this represents the truncation value of the IMF, not the maximum mass in Serpens B.}
\end{center}
\caption{
 \label{tab:photoclusts} Table of cluster and association properties used to generate a model cluster environments. Above the line are those for which properties are taken directly from \citet{Zwa10}. Below the line properties are found in independent sources (see Appendix \ref{sec:clust_models}). }
  \end{minipage}
\end{table*}

We adopt the following approach in modelling real clusters. First we choose clusters for which there exist consistent measurements of the half-mass and core radius. To construct a cluster we fit stellar positions consistent with the \citet{Els87} surface density profile:
\begin{equation}
\label{eq:surf_model}
\Sigma (d_\mathrm{c}) = \Sigma_0 \left( 1+\frac{d_\mathrm{c}^2}{a^2} \right)^{-\frac{\gamma}{2}} 
\end{equation} as a function of the projected distance $d_\mathrm{c}$ from the cluster centre, where $a$ is a scale parameter such that $r_\mathrm{core}$ is the distance at which the \textit{surface} density drops to half of its central value:
$$
r_\mathrm{core} = a \sqrt{(2^{2/\gamma} -1)}
$$ The associated volume density profile is
\begin{equation}
\label{eq:rho_model}
\rho(r_\mathrm{c}) = \rho_0 \left( 1 + \frac{r_\mathrm{c}^2}{a^2} \right)^{-\frac{(\gamma+1)}{2}}
\end{equation} where 
$$
\rho_0 = {\int_0^\infty y\Sigma(y)  \, \mathrm{d} y} \bigg/{ \int_0^\infty 2 z^2 \left( 1+\frac{z^2}{a^2} \right)^{-\frac{(\gamma+1)}{2}} \, \mathrm{d} z}.
$$ 
Where it is not defined in the literature, the value of $\gamma$ is obtained by fitting to $r_\mathrm{core}$ and the effective or half-light radius $r_\mathrm{eff}$:
$$
\int_0 ^{r_\mathrm{eff}}  y \left( 1+ \frac{y^2}{a^2} \right)^{-\gamma/2}  \, \mathrm{d}y = \frac{ M_\mathrm{clust}}{4\pi \Sigma_0} .
$$ In cases where $\gamma \leq 2$ we introduce a truncation radius $r_\mathrm{t}$ such as to give $r_\mathrm{eff}$ consistent with observations:
$$
\int_0^{r_\mathrm{t}} y\left(1+\frac{y^2}{a} \right)^{-\gamma/2} \,  \mathrm{d}y = 2\int_0^{r_\mathrm{eff}} y\left(1+\frac{y^2}{a} \right)^{-\gamma/2} \,  \mathrm{d}y.
$$  Hence we obtain a volume density profile $\rho$ as a function of radius within the cluster $r_\mathrm{c}$.  We note that introducing a truncation radius means that the 2D profile deviates slightly from the fitted profile for large $r_\mathrm{c}$ when truncating the 3D profile. 

To obtain a number density in terms of radius $r_\mathrm{c}$ the mass density is divided by the average stellar mass obtained from the initial mass function \citep[IMF][]{Kro93}:
\begin{equation}
\label{eq:imf}
  \xi(m)\propto \begin{cases}
               m^{-1.3} \quad \mathrm{for } \, 0.08 \, M_\odot \leq m < 0.5 \, M_\odot\\ 
              m^{-2.2} \quad \mathrm{for } \, 0.5 \, M_\odot \leq m < 1.0 \, M_\odot\\
              m^{-2.7} \quad \mathrm{for } \, 1.0 \, M_\odot \leq m < 100 \,M_\odot \\
              0 \qquad \quad \, \mathrm{else}
            \end{cases}
\end{equation} such that $\xi$ is normalised and continuous (although a slightly different IMF is used in the case of Cygnus OB2, see Appendix \ref{sec:cygmodel}).  However, as the FUV flux is sensitive to the mass of the most massive star in the cluster, we need to truncate the IMF above the chosen $m_\mathrm{max}$. To choose this mass, we note that \citet{Mas08} find that $m_\mathrm{max}$ is consistent with random drawing for clusters with a given number of stellar components $N_\mathrm{clust}$. We therefore draw the $m_\mathrm{max}$ distribution from our IMF, Equation \ref{eq:imf}, the results of which are shown in Figure \ref{fig:pmmax}. Our stellar atmosphere models are limited to stellar masses $<100M_\odot$ (Section \ref{sec:masslum}), which is therefore our upper limit on $m_\mathrm{max}$. A posteriori we will find that photoevaporation dominates over tidal truncation. To confirm this result as unambiguously as possible we seek to underestimate the influence of the FUV flux on a PPD population where there exists uncertainty in the correct prescription. For this reason, where we do not have an observational value for the most massive star in a cluster, we choose $m_\mathrm{max}$ one standard deviation below the median to give an underestimate of the photoevaporation rate. For example, in the case of the ONC, the most massive star is a component of a binary, $\theta^1 $ Orionis C, with mass $\sim 37\, M_\odot$ \citep{Kra09} which is our adopted $m_\mathrm{max}= m^\mathrm{obs}_\mathrm{max} = 37 \, M_\odot$. In the case of NGC 3603 we do not have an observed maximum stellar mass, and therefore adopt the conservative estimate $m_\mathrm{max} = m^{-\sigma}_\mathrm{max}= 67 \, M_\odot$. The adopted properties for clusters are summarised in Table \ref{tab:photoclusts}. The first six regions come directly from \citet{Zwa10}. Other specific environments for which we have taken data from other sources are discussed in Appendix \ref{sec:clust_models}.

\subsection{UV Luminosity and Stellar Mass}
\label{sec:masslum}
\begin{figure}
	\includegraphics[width=\columnwidth]{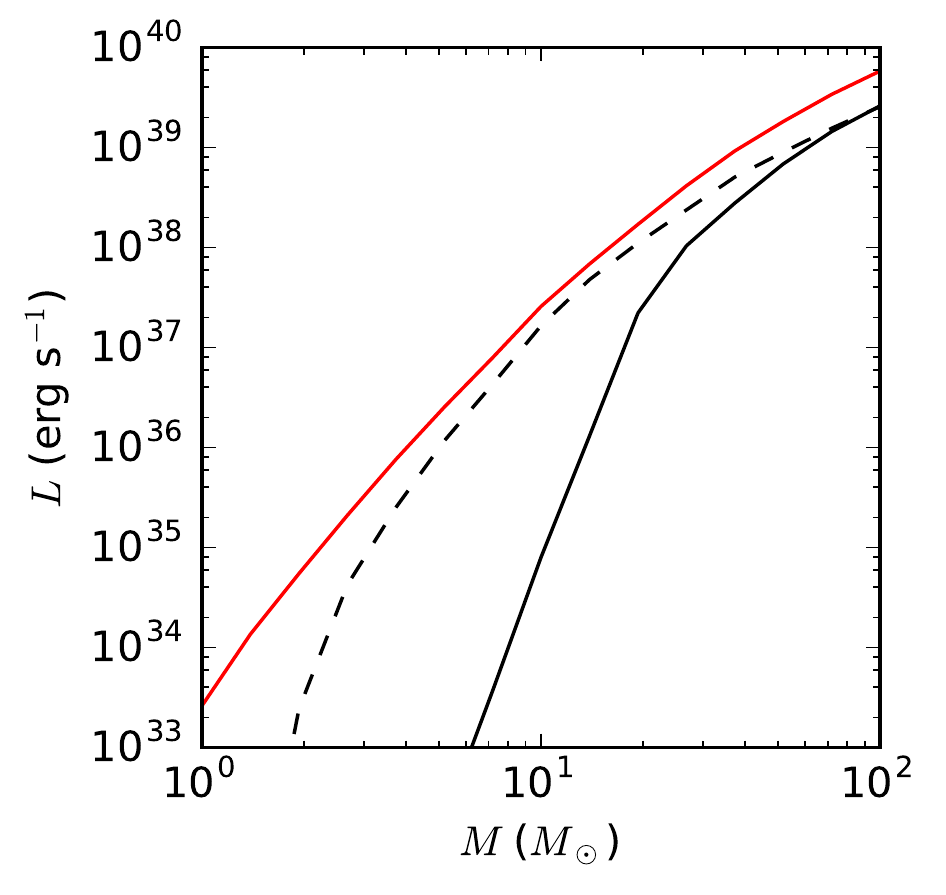}
    \caption{Stellar luminosity as a function of mass based on the models of \citet{Sch92} and \citet{Cas04}, which can be compared with the results of \citet{Arm00}. The red line indicates the total luminosity, while the black dashed and solid lines represent the FUV and EUV luminosities respectively.}
    \label{fig:L_mstar}
\end{figure}

To calculate the UV flux for a star of a given mass we follow the same method as \citet{Arm00} for stars with a mass in the range $1- 100 \, M_\odot$. The total luminosities and effective temperatures $T_\mathrm{eff}$ are taken from the stellar model grids of \citet{Sch92}, using the results for $Z=0.02$ and the output closest to the time $1$~Myr. These are combined with the stellar atmosphere models by \citet{Cas04} to give the wavelength dependent luminosity.

The FUV photons have energies in the range $6$~eV$<h\nu<13.6$~eV, while photons with energies higher than $13.6$~eV are considered extreme ultraviolet (EUV). The results shown in Figure \ref{fig:L_mstar}, which are in agreement with those of \citet{Arm00}, despite our use of the more recent atmosphere models. We can now apply these results to establish an external flux contribution for a given member of a cluster due to all other members with mass $1\, M_\odot < m < 100 M_\odot$.

\subsection{Local Environment Distribution}
\label{sec:local_environs}
\begin{figure*}
	\includegraphics[width=\textwidth]{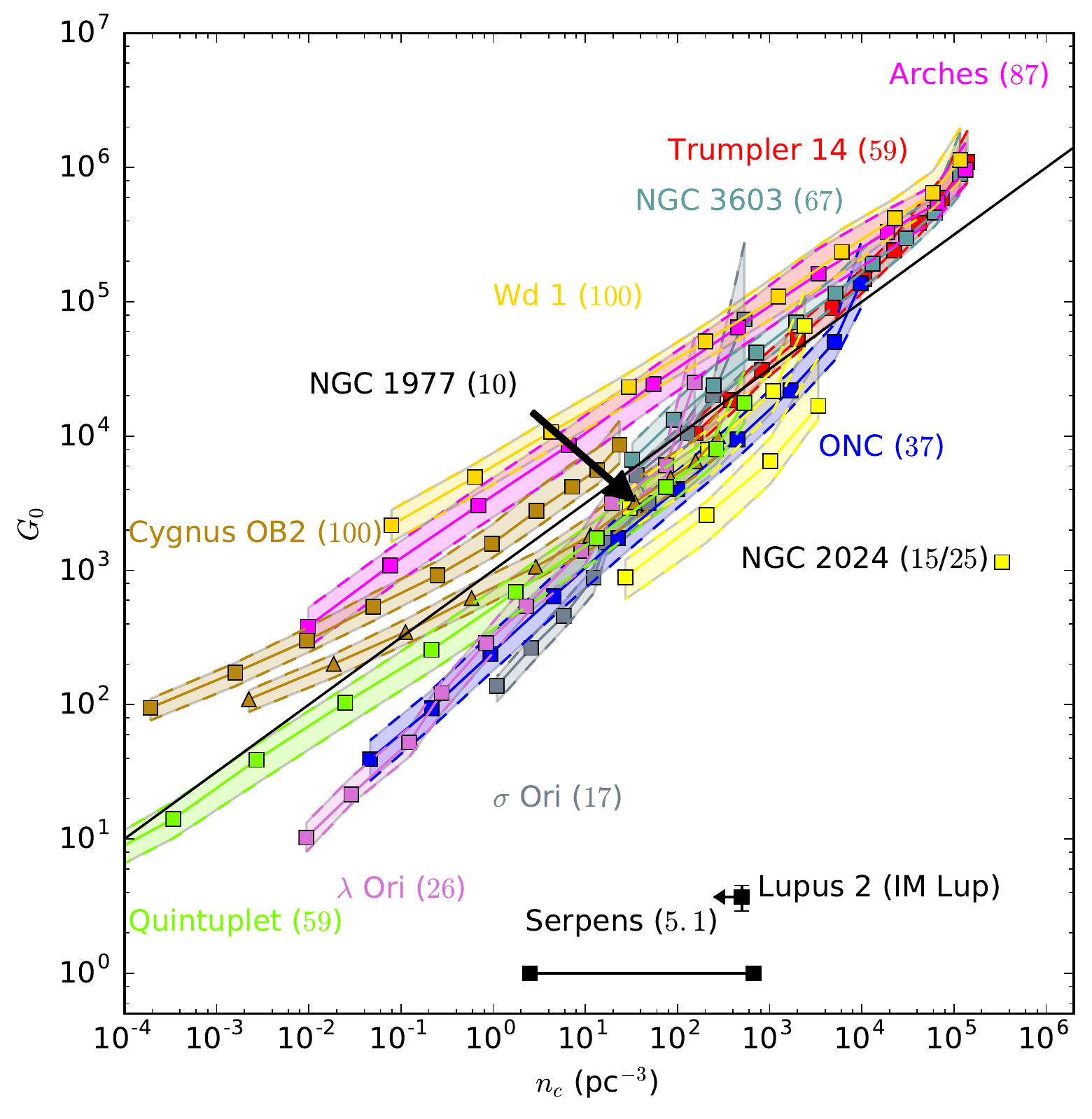}
    \caption{Contours follow the local number density and FUV flux within each cluster. All clusters are divided into radial bins and the mean flux and number density in that bin are represented by the square markers, except in the case of a contour for Cygnus OB2 marked by triangles which are the results when sub-structure is considered. The shaded regions represent the standard deviation ($\pm 1\sigma$) of the flux in each radial bin. The numbers in brackets represent the assumed maximum stellar mass in solar masses for each cluster. The solid black line follows $G_0 = 10^3 (n_\mathrm{c}/\mathrm{pc}^{-3})^{1/2}$.}
    \label{fig:nvsf}
\end{figure*}

In Figure \ref{fig:nvsf} the results of our cluster modelling are shown. The contours follow the density profile of the cluster, where stars are binned by radius. This yields a mapping between number density and FUV flux (assuming each cluster model is spherically symmetric). In some cases we have not directly modelled the clusters (Lupus, Serpens, and NGC 1977 - see Appendix \ref{sec:clust_models} for details). In particular we note that the Cygnus OB2 results altered to match the density and flux distribution of \citet{Gua16} are shown with triangular markers in Figure \ref{fig:nvsf}.

We find that for relatively massive clusters, $M_\mathrm{clust} \gtrsim 10^3\, M_\odot$, there is comparatively little dispersion in FUV flux for a given local number density. The relationship 
\begin{equation}
\label{eq:nvg0}
G_0 = 1000 \left(\frac{n_\mathrm{c}}{\mathrm{pc}^{-3}} \right)^{1/2}
\end{equation} is shown as a solid black line in Figure \ref{fig:nvsf}. It describes the contours for the massive clusters within a factor $\sim 3$, irrespective of $m_\mathrm{max}$, although we do not investigate cases for which $m_\mathrm{max}>100 \, M_\odot$ here. The gradient of the individual cluster contours in $n_\mathrm{c}-G_0$ space is dependent on the radial profile of the stellar density. In regions where $n_\mathrm{c}$ falls steeply with radius, the flux increases less rapidly with $n_\mathrm{c}$. This is expected given the reduced distance of the stars at a low $n_\mathrm{c}$ from the centre of the cluster for steep density profiles.

The fact that most clusters follow this relationship is a different realisation of the results of \citet{Arm00}, in which total FUV flux expected to be contributed as a function of stellar mass (considering a realistic IMF) becomes much less steep above $m \gtrsim 40 \, M_\odot$. This is due to the flattening of the luminosity as a function of stellar mass in this wavelength range. We note that this is not the case for the EUV flux, which also has a truncating influence on a PPD distribution but we do not address here.

We hold the physical discussion of the environments depicted in Figure \ref{fig:nvsf} until we have reviewed the photoevaporation and tidal truncation rate physics for various $G_0$, $n_\mathrm{c}$. From this analysis we will deduce where we might expect each truncation mechanism to dominate. For the reader who is only interested in the results of our analysis, we re-address Figure \ref{fig:nvsf} in this context in Section \ref{sec:comparison}. Appendix \ref{sec:specdisc} contains some discussion about the assumptions made in producing the contours for specific clusters for which there were modelling complications.


\section{Tidally Truncated Disc Radii} 
\label{sec:tidal}

As discussed in the introduction, in order to calculate disc radius evolution in a stellar cluster, the theoretical treatment available in the literature for the post-encounter disc radii requires updating. Previous works have made parametrisations of the truncation radius due to star-disc encounter, for example \citet{Bre14} find an empirical relation for the truncation radius of a disc of test particles for a range of perturber to host mass ratios $M_2/M_1$. However, as with the previous investigation of \citet{Hal96}, this calculation was not performed over an exhaustive range of disc orientations such that angle averaged results could be obtained. Nor were hyperbolic trajectories considered. Clearly a prescription for the former is necessary to apply to general encounters in a cluster. It also turns out that many encounters which occur in a cluster with a realistic distribution of stellar masses are highly eccentric \citep[e.g.][]{Vin16}, and therefore an evaluation of the influence of hyperbolic encounters on a disc is also required. 

We note that the recent study by \citet{Bha16} attempted to expand on \citet{Bre14} by angle averaging over disc truncation radii. However the fitted prescription for the post-encounter disc radius is not scale free since it would imply that the ratio of post-encounter radius to pericentre distance would depend on the absolute value of the latter. For this reason, while we still expand on the parameter space by considering different eccentricity encounters, we do not make assumptions about the form of our solution based on \citet{Bha16}, and use a fresh approach for finding the mass dependence for angle-averaged tidal truncation radii.

Therefore in this Section we first build on previously developed models for tidal truncation for an arbitrary encounters. From Section \ref{sec:secondpart} onwards we apply our model to cluster environments by considering a theoretical encounter rates in order to contextualise our findings, the results of which are discussed in Section \ref{sec:cluster_evol_results} and beyond.

For the disc evolution we consider the case both of a solar mass star, and smaller stars at the hydrogen burning limit. In all cases we take a canonical initial outer disc radius of $100$~au and apply statistical arguments to follow the radius evolution over $3$~Myr of evolution for a range of cluster densities and velocity dispersions. There is observational evidence that disc around brown dwarves are more compact than around solar mass stars \citep{Alv13,Tes16, Taz17,Tri17}. By studying the differential effect of mass on expected disc radius we aim to establish whether close encounters are a plausible mechanism for this difference.

\subsection{Numerical Method}
\label{sec:num_method}
We follow the same numerical method as in \citet{Win18} to evaluate the effect of a stellar encounter on a ring of test particles around a host star, which we review briefly here. The general Bulirsch-Stoer algorithm of the \textsc{Mercury} orbital integrator for solar-system dynamics is used \citep{Cha99}.

We have modelled each ring with $N=200$ particles, this being a compromise between computational expense and accuracy (this choice is discussed in Appendix \ref{sec:numconv}). Such a ring of $N$ particles is then fixed at some distance $r$ from a central star of mass $M_1$. A second star of mass $M_2$ is placed on a trajectory at a time $50$ test particle orbits prior to closest approach, and integrated for the same time subsequent to that approach. While for different $r$ this does not physically correspond to the same phase difference the results are found to be insensitive to the initial location of the perturber.


We define two angles of orientation: the angle between the direction of pericentre and the line of intersection of the disc and the orbital plane, $\alpha$, and the angle between the angular momentum vector of the disc and that of the orbit, $\beta$ \citep[see][]{Ost94, Win18}. We will angle-average our solutions (Section \ref{sec:angavg_models}) so the precise definition of the disc orientation is not important to our results. 

\subsection{Outer Radius Definition}
\label{sec:rout_def}
In order to establish the outer disc radius after an encounter $R'_\mathrm{out}$, previous studies have established definitions based on some limit on the surface density of a disc \citep[e.g.][]{Bre14}. Here we define $R'_\mathrm{out}$ by the post-encounter circularisation of particles that remain bound after the encounter. Assuming Keplerian motion we have angular momentum $L\propto r^{1/2}$, and therefore the fractional change in radius for a particle $i$:

$$
\frac{\Delta r_i}{r} = \left( \frac{\Delta L_i}{L} \right)^2 + \frac{2 \Delta L_i}{L}
$$

We average the circularisation radii for all particles in the ring that remain bound and define the disc outer radius as being the maximum value of $r'$ for all the rings in the disc. We however add the requirement that only rings where $> 90\%$ of the particles remain bound after the encounter are used in determination of the new disc outer radius. If these rings are not excluded, the trajectories of a small number of particles introduce significant noise into the outer radius determination.

Hence the new outer radius of the disc is defined:
$$
R'_\mathrm{out} = \mathrm{max}\, \left\{ r+\Delta r : r < R_\mathrm{out} \, \, \, \mathrm{ and } \, \, \, N'/N>0.9\right\}
$$ and the change in outer radius is $\Delta R_\mathrm{out} =  R_\mathrm{out} - R'_\mathrm{out}$. In the case of close, coplanar, prograde and parabolic encounters, this definition yields the same truncation radius as in the literature $R'_\mathrm{out} \approx 0.28 (M_2/M_1)^{1/3} x_\mathrm{min}$ \citep[e.g.][]{Hal96, Bre14}.

\subsection{Modelling}
\label{sec:angavg_models}
To make our results applicable to general encounters, we aim to produce a set of equations to define the post-encounter outer radius $R'_\mathrm{out}$ as a function of the encounter parameters: the closest approach distance $x_\mathrm{min}$, the eccentricity $e_\mathrm{pert}$ and the ratio of the perturbing to host mass $M_2/M_1$. The orientation of the disc with respect to the perturbing star is also important for the truncation radius, however in order to simplify the models we address the angle-averaged results. These are given by 

$$
\left\langle \frac{\Delta R_\mathrm{out}}{R_\mathrm{out}} \right\rangle = \frac 1 {4\pi} \int_0^{2\pi} \mathrm{d} \alpha \int_0^{\pi}\mathrm{d}\beta \, \sin\beta\frac{\Delta R_\mathrm{out}}{R_\mathrm{out}}(\alpha, \beta) 
$$ where $\alpha$ and $\beta$ parametrise the disc orientation as described in Section \ref{sec:num_method}. We note that the angle averaging we perform is in fact the sum over trapezia using the outer radius results at intervals of $30^\circ$ in order to make calculations over the required range of angles computationally practicable. 

The nature of the fitting formula with which we model our results is discussed in Appendix \ref{sec:fitform}. In short, we model three distinct regimes in the closest approach distance $x_\mathrm{min}$ using six fitting parameters $\phi_{i=1,2,3}, \psi_{i=1,2,3}$. In the closest regime the disc radius is considered to be a fixed fraction of the closest approach for given mass ratio $M_2/M_1$ and eccentricity $e_\mathrm{pert}$. In the distant regime we assume the disc radius is unchanged. The model is highly simplified but we will find that it is sufficient in all the physically relevant regions of parameter space.

\subsection{Post-Encounter Disc Radius}

\label{sec:ringres}
\begin{table}
\begin{center}
 \begin{tabular}{c | c c c} 
 \hline
 Parameter  &  Value & $\sigma_+$ & $\sigma_-$ \\ [0.5ex] 
 \hline\hline
 $\phi_1$ & $0.629$ & $0.633$ & $0.624$ \\  
 $\phi_2$ & $0.112$  &  $0.114$ &  $0.109$ \\
 $\phi_3$ & $0.133$ & - &  -  \\
 $\psi_1$ & $0.301$ & $0.307$ & $0.296$  \\
 $\psi_2$ & $0.936$ & $0.947$ & $0.924$ \\
 $\psi_3$ &  $0.320$ & $0.323$ & $0.317$ \\[1ex] 
 \hline
\end{tabular}
\caption{ Fitting parameters and errors for our general model for post-encounter disc radius. All values are established using an MCMC implementation except for $\phi_3$. }
 \label{tab:fittingparams}
\end{center}
\end{table}

\begin{figure*}
	\includegraphics[width=0.95\textwidth]{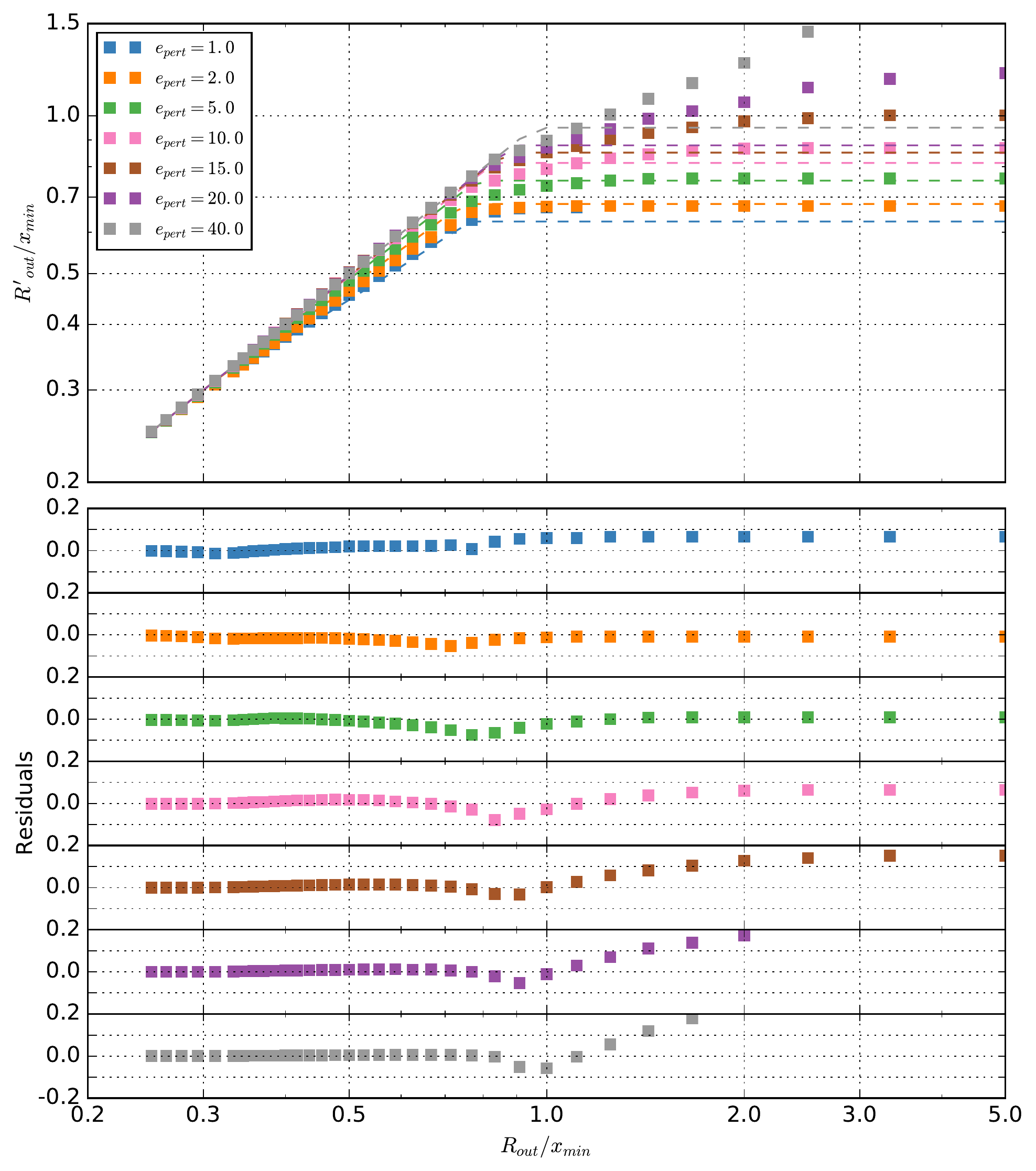}
    \caption{The angle-averaged post-encounter radius $R'_\mathrm{out}$ of a disc with initial radius $R_\mathrm{out}$ as a fraction of the closest approach distance of an encounter $x_\mathrm{min}$ where stellar components are of equal mass $M_2/M_1=1$. Simulation data points are shown as squares. The model, which is fitted to the data points where the perturber eccentricity $e_\mathrm{pert}\leq 20$, is shown by the dashed lines (see the text for details). The residuals are shown in the bottom panel.}
    \label{fig:model1}
\end{figure*}

Our model, the form of which is described by Equations \ref{eq:fdef} through \ref{eq:fullmod}, is fitted with the parameters summarised in Table \ref{tab:fittingparams}. Figure \ref{fig:model1} shows the results in the $M_2/M_1=1$ case. We find good agreement with the simulations within $10\%$ except in the limit of large $e_\mathrm{pert}$ and $R_\mathrm{out}/x_\mathrm{min}$ (penetrating, hyperbolic encounters). Encounters in this region of parameter space are both unlikely and expected to yield capture scenarios and PPD destruction. However, we note that caution should be used when applying our results for arbitrary masses and eccentricities. Due to the difficulty with the highly hyperbolic case, we exclude the $e_\mathrm{pert} = 40$ results during our fitting procedure. 

\begin{figure*}
	\includegraphics[width=0.95\textwidth]{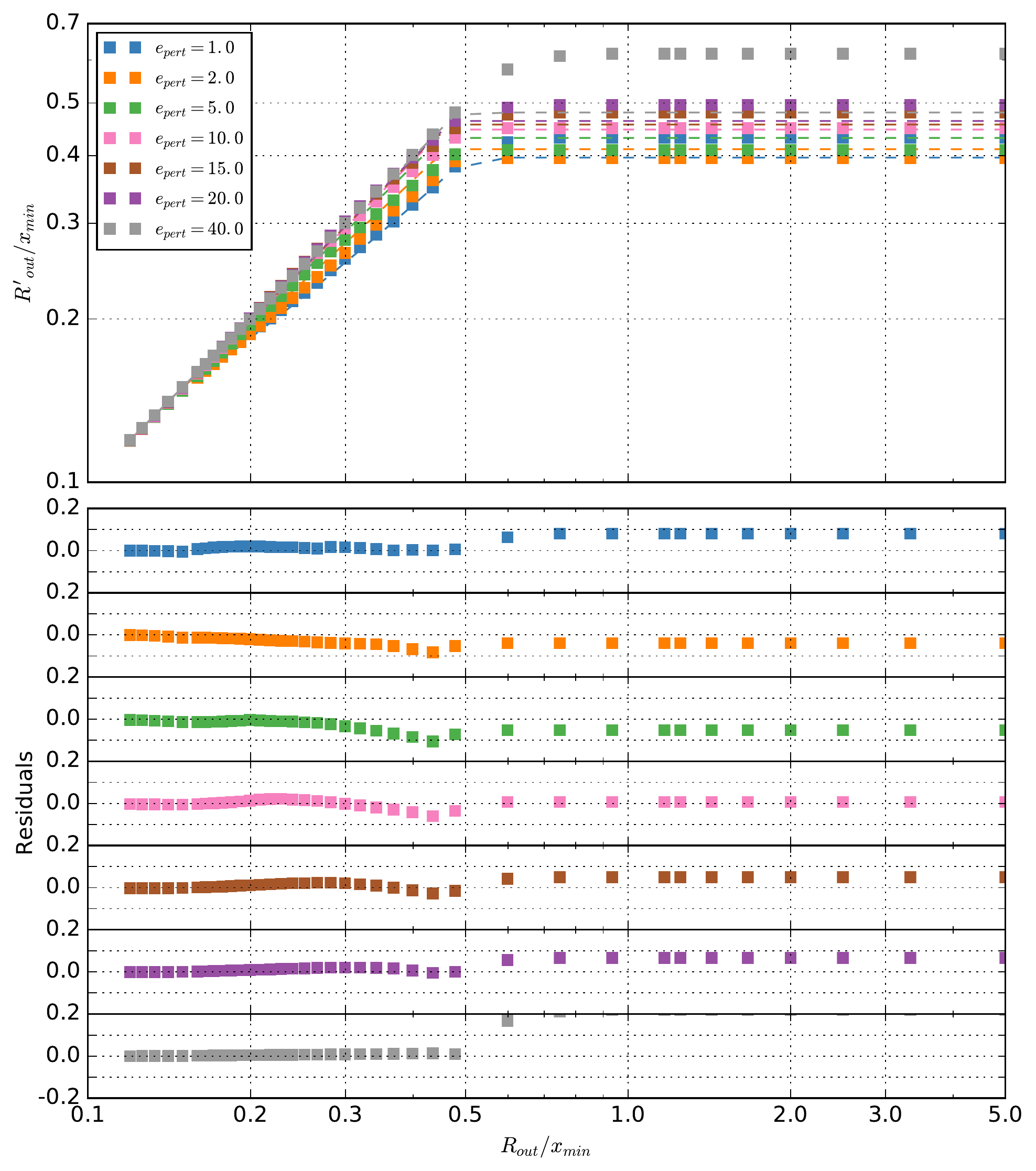}
    \caption{As in Figure \ref{fig:model1} except the ratio of the perturber to host stellar mass is $M_2/M_1=10$. The model values (dashed lines) are fitted only to the simulation data of the $M_2/M_1=1$ case except in the asymptotic limit $R_\mathrm{out}/x_\mathrm{min} \gg 1$, where an additional mass dependent factor is fitted (see text for details).}
    \label{fig:model100}
\end{figure*}

In order to obtain the final fitting parameter $\phi_3$, we choose the value which best fits the simulation results for $M_2/M_1=10$, shown in Figure \ref{fig:model100}. The form of the fitting function ensures that $R'_\mathrm{out}$ is less dependent on $e_\mathrm{pert}$ for large $M_2/M_1$, which we find is partially true. In reality, the relationship between these parameters is more complex, and in the $M_2/M_1=10$ case we see that $R'_\mathrm{out}/x_\mathrm{min}$ does not vary monotonically with $e_\mathrm{pert}$. Our prescription is only out by more than $20\%$ in the extremely hyperbolic case $e_\mathrm{pert}=40$, and for all the rest of the results the model is accurate within $10\%$.

We further test our model for $M_2/M_1=100$, where the dependence of the truncation radius on $e_\mathrm{pert}$ is more complex and difficult to model accurately than at lower mass ratios. Despite this, the majority of our numerical results remain within $\sim 20\%$ of the model predictions. Given that penetrating encounters with a mass ratio $M_2/M_1 \sim 100$ occur with low probability given the form of the IMF and typical velocity dispersions in clusters, we do not address a more sophisticated treatment of this region of parameter space here.

\begin{figure*}
	\includegraphics[width=0.95\textwidth]{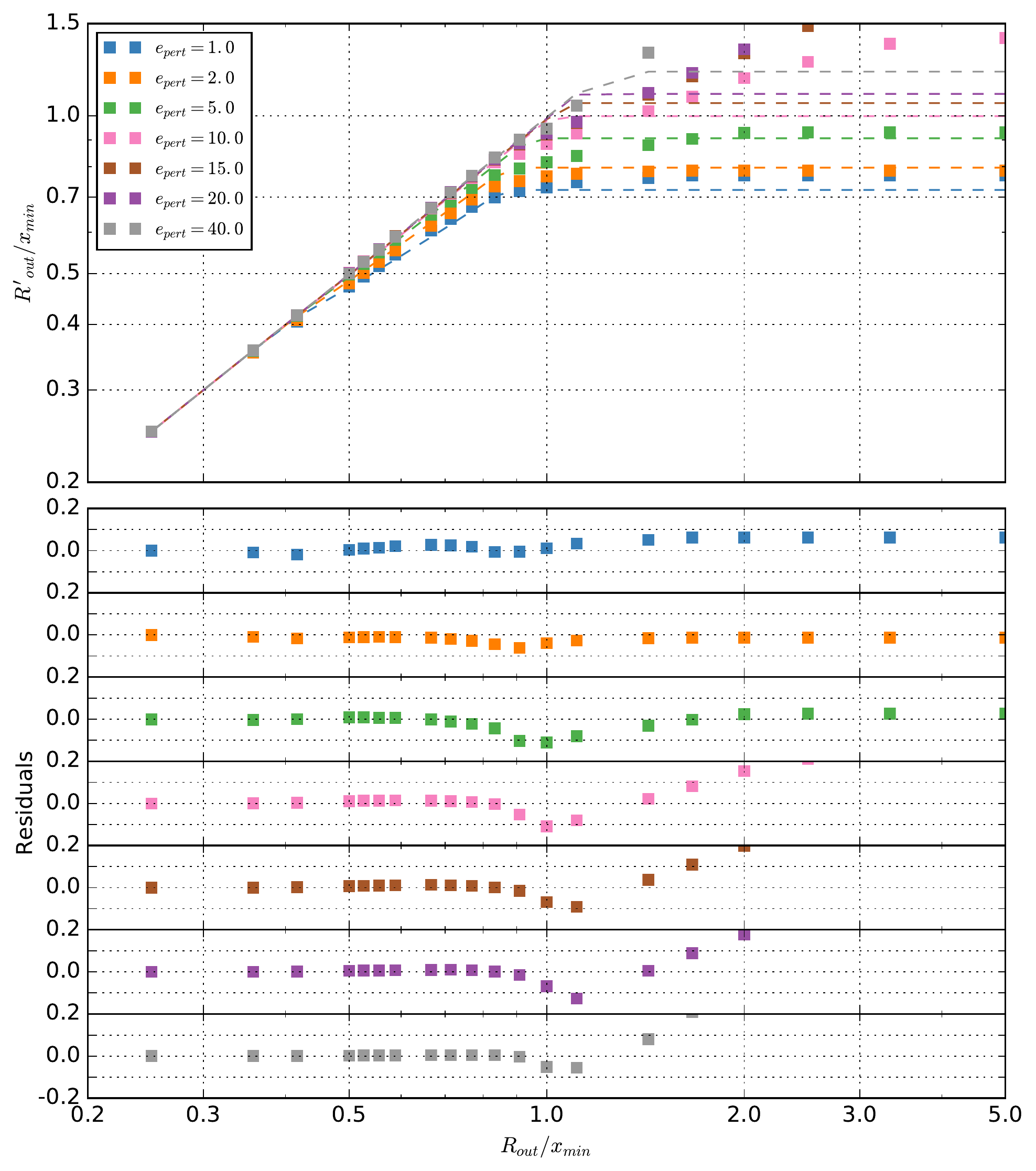}
    \caption{Angle-averaged outer radius of a disc due to an encounter with a star of varying closest approach distance $x_\mathrm{min}$ and trajectory eccentricity $e_\mathrm{pert}$. Model (dashed lines) and simulation results (squares) for the case where $M_2/M_1 =0.5$. }
    \label{fig:modelhalf}
\end{figure*}

A comparison between the simulation results and the model is shown for $M_2/M_1=0.5$ in Figure \ref{fig:modelhalf}. No further adjustment to the model parameters is applied in this case. Results are once again within $10\%$ of the model for $e_\mathrm{pert}< 10$, and the discrepancies largely occur in regions which are both unlikely (highly hyperbolic and close encounters) and prone to inducing binaries or disc destruction.

Additionally we investigate the effect of varying the particle number threshold for $N'/N$, and find that reducing it only influences the results for low probability (i.e. highly hyperbolic, penetrating) encounters. Our model remains in agreement with the simulation results within $10\%$ in the regions of parameter space which are of interest.

\subsection{Encounter Rate}
\label{sec:secondpart}

\label{sec:clustmod}
\begin{figure}
	\includegraphics[width=\columnwidth]{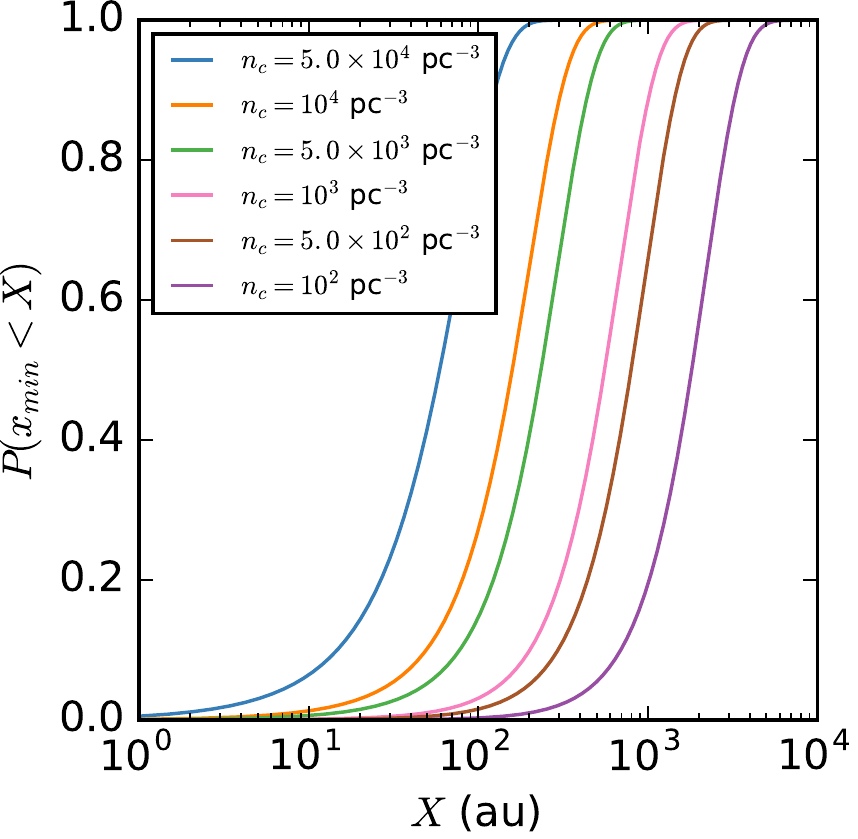}
    \caption{The probability of a star having an encounter for which $x_\mathrm{min} < X$ in different stellar densities after 3 Myr. The cluster is assumed to have uniform density and be composed of stars with mass $1 \, M_\odot$, with one dimensional velocity dispersion $\sigma_v =4$~km/s. }
    \label{fig:encstheory}
\end{figure}
 
In order to generate appropriate cluster models, we must first establish the encounter rate for varying environments. This is discussed in Appendix \ref{sec:encrate}, and is dependent on the IMF, the stellar number density $n_\mathrm{c}$, the velocity dispersion $\sigma_v$ and sub-structure \citep[e.g.][]{Olc06, Cra13}. The latter can be approximately parametrised by employing two additional quantities; the total number of stars $N_\mathrm{c}$ and the initial fractal dimension $D_0$. The differential encounter rate is denoted $\mathrm{d}\Gamma = \gamma \, \mathrm{d}x_\mathrm{min} \, \mathrm{d}V^2 \, \mathrm{d}M_2$ and defined in Equation \ref{eq:encrate}. In the most general form we want to estimate the probability that an encounter occurs in a small region of parameter space: its closest approach in a spatial range $\delta x_\mathrm{min}$ around $x_\mathrm{min}$, a perturbing mass range $\delta M_2$ around $M_2$, the range of dimensionless square relative velocity at infinity $\delta V^2$ and a time range $\delta t$. For convenience we label such a box $A$, its volume $\delta A = \delta x_\mathrm{min} \, \delta V^2 \, \delta M_2 \, \delta t$ and a coordinate in parameter space $a$. Assuming that encounters are uncorrelated, they can be modelled as a Poisson process and thus the probability that an encounter will occur in $A$ is approximately
\begin{equation}
\label{eq:probbox}
P(a \in A) \approx 1 - \exp\left(- \gamma \, \delta A \right)
\end{equation} where $\gamma$ is evaluated at some point in $A$. In the limit $\delta A \rightarrow 0$ the term in the exponent can be integrated such that the probability of an encounter in any given range can be calculated. 

Given this general encounter rate, we identify six distinct cluster models to investigate. For the most simplifying conditions we consider a cluster comprised of equal mass stars without any sub-structure. Two additional models are required to examine the effect of a realistic IMF and time-dependent sub-structure. In each case a `high-density' and `low-density' model give a sense of the dependence of disc evolution on $n_\mathrm{c}$. Practically, we can use the theoretical encounter rates to estimate what high- and low-density regions are of interest. Equation \ref{eq:probbox} is evaluated as an integral over $3$~Myr for a cluster comprised of solar mass stars and without sub-structure in Figure \ref{fig:encstheory}. We choose a high-density model with $n_\mathrm{c} = 5 \times 10^4$~pc$^{-3}$ for which a significant fraction undergo an encounter such that $x_\mathrm{min} \lesssim 100$~au.

For a `low-density' model we choose $n_\mathrm{c} = 10^4$~pc$^{-3}$. We note that this is not low-density in that it is higher than typical densities suggested by local observations, although theoretically making statistical predictions on such properties is strongly dependent on the formation environment \citep{Kru12}. \citet{Bre10} found that the stellar surface density distribution in the local $500$~pc varies up to $\sim 10^3$~pc$^{-2}$, with a peak (by population) at $\sim 22$~pc$^{-2}$. However, the majority of stars in a region with $n_\mathrm{c} = 10^4$~pc$^{-3}$ have closest encounters such that $100$~au $ < x_\mathrm{min} < 10^3$~au. Hence this represents an intermediate environment, approximately the lowest density where we expect a population of PPDs to undergo any significant tidal truncation.

We additionally need to define a 1D velocity dispersion $\sigma_v$. In many cases interpreting real cluster properties is not straightforward. For example, many clusters appear to be super-virial, and it is possible that this is because velocity dispersions are overestimated due to binaries \citep{Gie10}. Incompleteness and uncertainties in establishing cluster membership also contribute to uncertainties in local stellar densities. For more detailed discussion see \citet{Sto10}. For a review of the properties of young massive clusters see \citet{Zwa10}. We assume a velocity dispersion of $\sim 1-5$~km/s is usual in most clusters \citep[e.g.][]{Hil98, Cla05, Roc10, Cla12}, although in some clusters a larger $\sigma_v$ is observed \citep[such as Cygnus OB2][]{Wri16}. In our models we initially assume $\sigma_v= 4$~km/s, but subsequently examine the effect of varying this value.

For the cluster models including an IMF we use $\xi$ as in Equation \ref{eq:imf} \citep{Kro93}. In models for a single stellar mass, all stars are assumed to have $m=1M_\odot$. All the cluster models are summarised in Table \ref{tab:clustmodels}. These models are not intended to be a realistic representation of an overall cluster, not least because in the dynamic evolution of a real cluster the stellar density and mass distribution is likely to be spatially dependent. They are instead intended to reproduce the local conditions and therefore apply to a disc which has spent its life in a fixed stellar environment.
 
\begin{table}
\begin{center}
 \begin{tabular}{c | c c c c c } 
 \hline
 Model  & $n_\mathrm{c}$ (pc$^{-3}$) & $D_0$ & $\sigma_v$ (km/s) & $N_\mathrm{c}$ & IMF\\ [0.5ex] 
 \hline\hline
 A & $10^4$ & 3.0 & 4.0 & -  & - \\  
 B & $5\times 10^4$ & 3.0 & 4.0 & -   & -\\ 
 C & $10^4$ & 2.0 & 4.0 & $10^3$ & -\\ 
 D & $5\times 10^4$ & 2.0 & 4.0 & $10^3$ & - \\
 E & $10^4$ & 3.0 & 4.0 & - & $\xi$ \\ 
 F & $5\times 10^4$ & 3.0 & 4.0 & - & $\xi$ \\ [1ex] 
 \hline
\end{tabular}
\caption{Summary of cluster model parameters. In cases where the fractal dimension $D_0 =3.0$, uniform conditions, the number of stars in the cluster is irrelevant. Where the IMF is not listed all stars are assumed to be of solar mass. }
 \label{tab:clustmodels}
\end{center}
\end{table}

\subsection{Numerical Method}
\label{sec:clust_numerics}

We adopt a {\it Monte Carlo} approach in quantifying the stochastic evolution of the outer radius of a disc embedded in a stellar cluster. For each model in Table \ref{tab:clustmodels}, $10^3$ disc evolutions are calculated. In the case that an IMF is included (Models E and F), the mass of the host star is drawn from the distribution defined by $\xi$ in Equation \ref{eq:imf}. 

For each disc, the parameter space (over time, mass of perturbing star and spatial separation) is divided into grid cells, each of size $\delta A$ as defined in Section \ref{sec:clustmod}. Each grid cell is assigned a random number $u\in [0,1)$. If $u<P(a\in A)$, as defined in Equation \ref{eq:probbox}, then an encounter is logged. The point in parameter space is then drawn at random from within the grid cell $A$.

Some consideration as to the maximum size of each grid cell $\delta A$ is required. The size of a partition for each variable $a_i$ should be limited such that $\delta a_i \ll  \gamma / \left| \nabla_{a_i} \gamma \right|$. For  the time dimension the size of $\delta t$ is important only in cases where sub-structure is included, when it is necessary that $\delta t \ll \tau_\mathrm{cross}$, the crossing time of the cluster. 
 
Finally, the probability of two encounters occurring in the same grid cell $ A$ should be small. If we take a maximum probability that two events occur in the same cell as $1\%$, this means that $P(a\in A)<0.1$. From Equation \ref{eq:probbox}, this gives
$$
\delta A < \frac{-\ln(0.9)}{\gamma}
$$ We therefore chose our grid cells carefully to adhere to these conditions, varying the cell partitions depending on the model parameters.

In this manner a series of encounters are assigned to a set of points $\{ a \}$ in parameter space. The encounters are then applied to the disc under consideration in chronological order, such that the disc response is appropriate for each sequential encounter. The initial outer radius is defined to be $100$~au regardless of host mass, and the outer radius of the disc responds to subsequent encounters as described in Section \ref{sec:ringres}.

\subsection{Cluster Evolution Results}

\label{sec:cluster_evol_results}

\subsubsection{Uniform Density Cluster}

\begin{figure*}
	\includegraphics[width=0.95\textwidth]{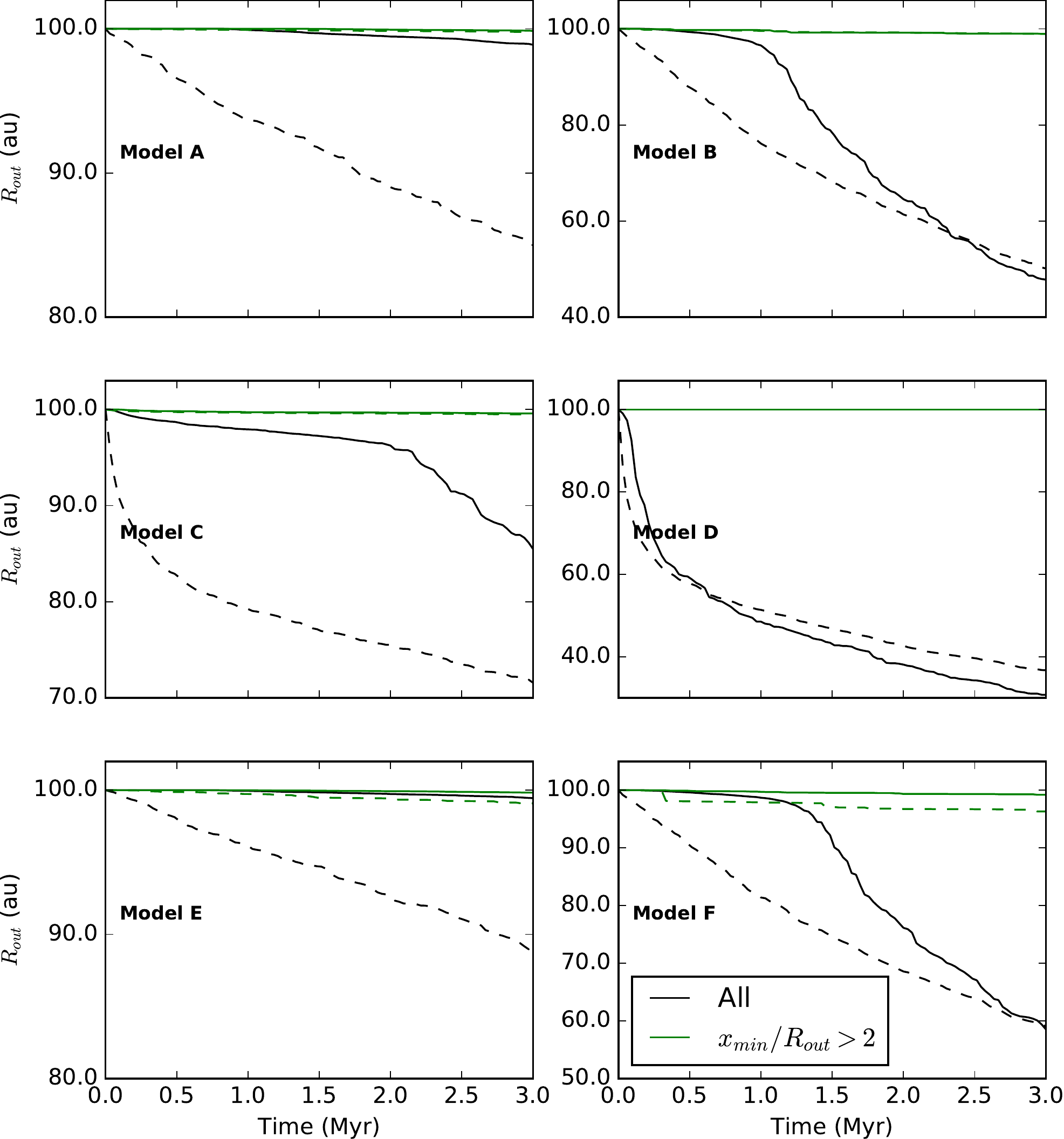}
    \caption{Median (solid lines) and mean (dashed lines) outer disc radius evolution for each cluster model. The black lines are results for all discs, the green lines are for discs which did not have any encounters such that $x_\mathrm{min}/R_\mathrm{out}< 2$. The parameters for each of the models are shown in Table \ref{tab:clustmodels}. }
    \label{fig:routevol}
\end{figure*} 

The outer radius evolution results for all the models are shown in Figure \ref{fig:routevol}, where the uniform density clusters composed of $1M_\odot$ stars are top left and top right, which are the low- and high-density cases respectively. In order to interpret these results physically, we have categorised them according to the closest encounter distance of the disc in question. The threshold $x_\mathrm{min}/R_\mathrm{out}<2$ is commonly taken as a criterion for significant disc truncation \citep[e.g.][]{Hal96}. We show in green the subset of discs that are only influenced by encounters such that $x_\mathrm{min}/R_\mathrm{out}>2$. The results including all discs are shown in black.

Clearly Model A yields no significant truncation, with the mean and median radii being little affected by encounters even when stars undergoing close encounters are included. The truncation extent is much greater in the high-density Model B, producing significantly reduced disc radii ($\sim 50$~au). More distant encounters still have little effect on the disc evolution, in agreement with \citet{Win18}. Further, the fraction of discs which do not have a close encounter is shown in Table \ref{tab:closeperc}, and we find that for such a high density cluster very few discs have only distant encounters over $3$~Myr.

\begin{table}
\begin{center}
 \begin{tabular}{c | c c c c c c} 
 \hline
 Model   & A & B & C & D & E & F \\ [0.5ex] 
 \hline \hline
 No close enc. & $32\%$ & $0.8\%$ & $11\%$ & $0\%$ & $40\%$ & $1.4\%$ \\
 \hline
\end{tabular}
\caption{The fraction of discs which did not undergo a close encounter (${x_\mathrm{min}}/{R_\mathrm{out}}< 2$) over $3$~Myr in each model.}
 \label{tab:closeperc}
\end{center}
\end{table}

\subsubsection{Structured Cluster}

For a cluster with sub-structure we expect to see a rapid evolution of disc outer radii at early times due to the effective stellar density enhancement (and therefore the cluster encounter rate), which is reduced over a crossing time $\tau_\mathrm{cross}$, as the cluster relaxes. This is confirmed in Figure \ref{fig:routevol}. In the cases that the cluster initially has sub-structure (Models C and D), $R_\mathrm{out}$ initially drops more rapidly. As the cluster ages however, the rate of change $\mathrm{d} R_\mathrm{out}/\mathrm{d}t$ decreases so that after $3$~Myr the average outer radii do not differ greatly from the unstructured case. 

We note that the extent of the difference between structured and unstructured models is dependent on the crossing time $\tau_\mathrm{cross}$, and therefore the number of local cluster members $N_\mathrm{c}$ for a given local stellar density. As we have fixed $N_\mathrm{c}=10^3$ in both of our sub-structured models, the cluster relaxes faster in the higher density Model D as opposed to the lower density Model C (see Appendix \ref{sec:encrate}). It is possible to increase the length of time for which the number density is enhanced by structure, but this pushes into regions of parameter space which are physically unlikely, requiring dense and large stellar populations. Similarly a smaller $\tau_\mathrm{cross}$ would reduce the time-scale over which density is enhanced.

\subsubsection{Cluster with Stellar Mass Distribution}

A realistic IMF is implemented in Models E and F, which are shown in the right and left bottom panels of Figure \ref{fig:routevol} respectively. The evolution of the disc radii for the global population is not significantly altered from Models A and B, without an IMF implementation. Some slight truncation for discs that only underwent distant encounters is observed in Model F due to the influence of high mass perturbers such that $M_2/M_1 \gg 1$. However, the fraction of discs which escape close encounters in this high density environment remains low at $1.4\%$.

While the overall statistical properties of the disc radii in the cluster are the same, we can make comparisons between stars of different mass within the cluster. We discuss the mass dependence of the truncation radii below.

\subsection{Mass Dependent Truncation}

\label{sec:mdeptt}

\begin{figure*}
	\includegraphics[width=\textwidth]{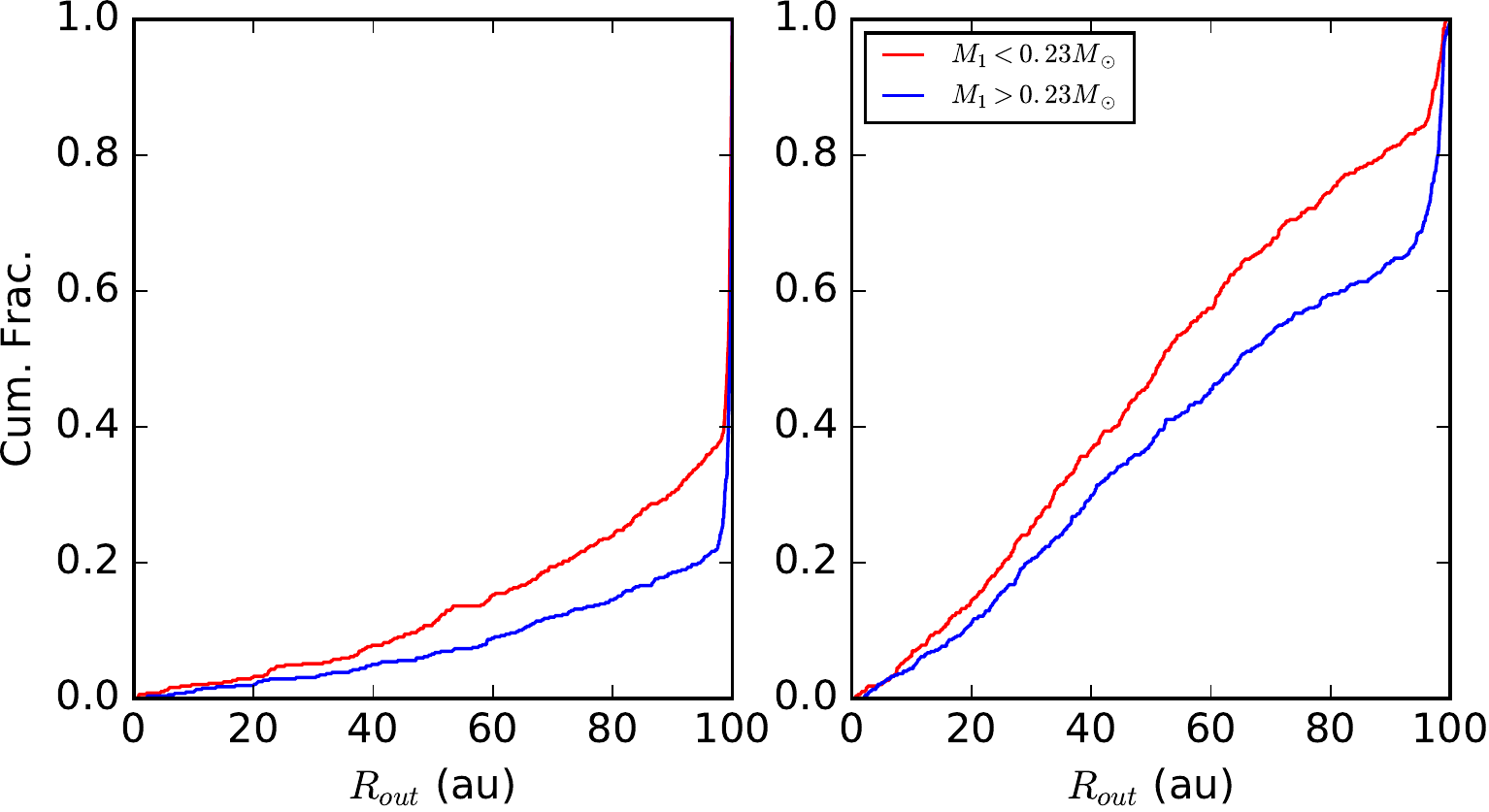}
    \caption{Cumulative fraction of the outer radius distribution $R_\mathrm{out}$ of discs evolving in a cluster wherein stellar masses are drawn from the IMF $\xi$ in Equation \ref{eq:imf} for Model E (left) and F (right). Samples of 1000 stars are divided into two approximately even samples by the host mass $M_1$ using the limit $0.23 \, M_\odot$.}
    \label{fig:mdist}
\end{figure*}

\begin{figure}
	\includegraphics[width=0.95\columnwidth]{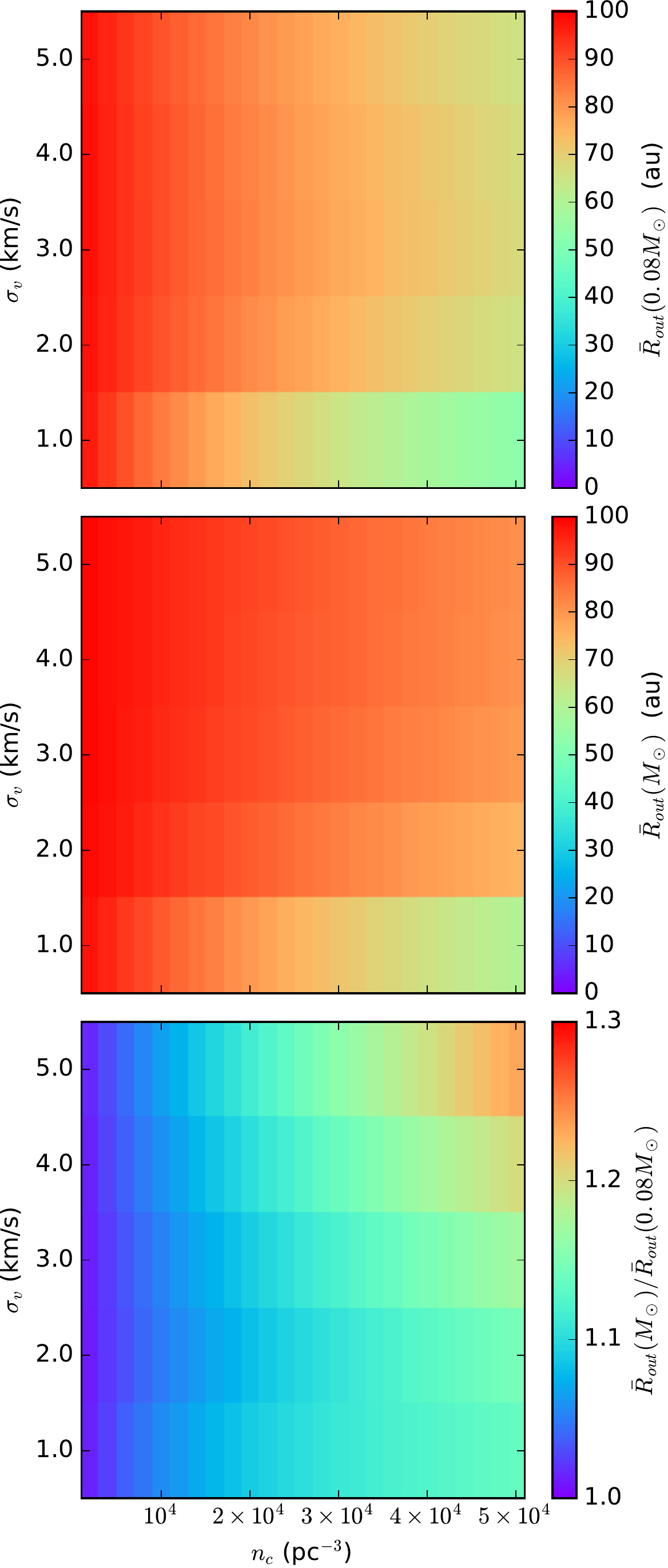}
    \caption{Top: mean outer radius of a disc hosted by a star of mass $0.08 \, M_\odot$  as a function of cluster properties, velocity dispersion $\sigma_v$ and number density $n_\mathrm{c}$. Middle: Mean outer radius for a disc around a $1 \, M_\odot$ star. Bottom: Ratio between the average disc outer radius of a disc hosted by a $1\, M_\odot$ and $0.08 \, M_\odot$ star. }
    \label{fig:r_m_ratio}
\end{figure}

Although we will find that tidal truncation is unlikely to ever drive disc radius distributions within a stellar population, we consider the dependence of tidal truncation effects on stellar mass for the sake of completeness. Given that a larger ratio $M_2/M_1$ yields greater angular momentum transfer, we should expect the final outer radius of a disc to increase with the host mass $M_1$. This is considered in Figure \ref{fig:mdist}. We divide the samples for Model E and F into two even subsets by a mass threshold ($0.23 \, M_\odot$) and plot the cumulative fraction of the outer radius distribution. Differences between the high- and low-mass sets are clear, and yield two-tail KS test $p$-values $\ll 0.05$ for our samples of $1000$ stars in both Models E and F, and they are still $\lesssim 0.05$ when the same analysis is considered for a random subset of $100$ stars. In the case that tidal truncation is the dominant truncation mechanism within a cluster, in principle it might be possible to find differences between the outer radius distributions in real observations. However putting constraints on disc radii to within $\sim 10\%$ for such a large sample of discs is realistically challenging. 


\subsection{Probability Averaging}
\label{sec:tidalavg}
In order to produce a more thorough exploration of the effect of changing $\sigma_v$ and $n_\mathrm{c}$, and given that final disc radii are largely determined by the `strongest' encounter, the average outer radius for a given stellar mass can be approximated by direct calculation in the following way. First we choose two comparison masses, $M^* = 0.08M_\odot$ and $1 \, M_\odot$, and for each $M^*$ we sort the regions of encounter parameter space $\{ A_i \}_{M^*}$ by how much a $100$~au disc is truncated, giving us a corresponding set of outer radii $\{R_{\mathrm{out, }i}\}_{M^*}$, ordered from smallest to largest. We then define a cumulative probability $C_i$, such that 
$$
C_{i+1} = C_i + P(a \in A_{i+1})\cdot (1-C_i)
$$ where $C_0 = 0$. Each $C_i$ is the probability that a stronger encounter than $A_{i+1}$ has occurred over the relevant time period. Hence the probability that $A_i$ is the strongest encounter is $P(a\in A_i)\cdot (1-C_{i-1})$. Then for a given host mass $M^*$, the mean outer radius can be approximated
$$
\bar{R}_\mathrm{out}(M^*) = \sum_{i=1} R_\mathrm{out}(A_i; M^*) P(a \in A_i; M^*)\cdot (1-C_{i-1}) 
$$ where $R_\mathrm{out}(A_i; M^*)$ is the post-encounter outer radius of a $100$~au disc hosted by a star of mass $M^*$ due to an encounter in the parameter range $A_i$. 

Applying this in varying cluster conditions yields the results in Figure \ref{fig:r_m_ratio}. These results are particularly interesting in the context of the outer radius dependence on the velocity dispersion within the cluster. Although overall encounter rates increase monotonically with $\sigma_v$, the likelihood of an encounter being hyperbolic also increases. In the case where $M_2/M_1 \lesssim 1$ the final outer radius of the disc is highly dependent on the eccentricity of the encounter. Therefore, particularly in the case of a relatively high mass star, tidal truncation is enhanced when the velocity dispersion within the cluster is small, even though there are fewer close encounters. Hence a prescription for the dependence of truncation extent on eccentricity is an important addition to the theory of star-disc encounters. We find that the most significant tidal truncation for all stellar masses is likely to occur in regions with $\sigma_v \lesssim 1$~km/s and $n_\mathrm{c} \gtrsim 2 \times 10^4$~pc$^{-3}$. 

The ratio of the average outer disc radii in different cluster conditions is plotted in the bottom panel of Figure \ref{fig:r_m_ratio}.  The ratios between the average radius for the $\sigma_v$, $n_\mathrm{c}$  chosen Models E and F, are found to be $1.06$ and $1.20$ respectively. The corresponding ratios obtained from the mean values of the subsets in Figure \ref{fig:mdist} are $1.06$ for Model E and $1.17$ for Model F. These values are comparable despite inclusion of a distribution of host masses in the latter case. Figure \ref{fig:r_m_ratio} can therefore be considered a valid comparison to real datasets over a range of host masses divided into subsets. Our results indicate that the region of cluster parameter space for which tidal truncation of PPDs is most significant (low-velocity, high-density) is not the same as the region of the greatest distinction between different stellar masses (high-velocity, high-density). Samples obtained from environments in which velocity dispersions are smaller than in Model E and Model F might exhibit less clear differences when analysed as in Figure \ref{fig:mdist}.

\subsection{Conclusions on Tidal Truncation}

We have presented a full investigation of PPD radius distributions driven by tidal encounters in local environments with varying stellar density and velocity dispersion. Our main conclusions are as follows:

\begin{itemize}
\item  The expected (mean) outer disc radius $\bar{R}_\mathrm{out}$ (from a initial outer radius of $100$~au) in a given environment is dependent on the local stellar number density $n_\mathrm{c}$, velocity dispersion $\sigma_v$ and host mass $M_1$. The value of $\bar{R}_\mathrm{out}$ is minimised for large $n_\mathrm{c}$, small $\sigma_v$ and small host mass $M_1$. 
\item We find that no environments for $n_\mathrm{c}<10^4$~pc$^{-3}$ yield $\bar{R}_\mathrm{out}<80$~au within $3$~Myr regardless of  $\sigma_v$ and $M_1$. In most cases $\bar{R}_\mathrm{out}$ is larger than this, almost unchanged from the initial value. At higher local number densities $n_\mathrm{c} \sim 5 \times 10^4$~pc$^{-3}$ we find that tidal encounters significantly truncate PPDs below $100$~au. We therefore adopt a fiducial density threshold above which tidal truncation becomes significant in PPD evolution $n_\mathrm{c} > 10^4$~pc$^{-3}$.
\item The differential effect of host mass $M_1$ on the outer radius distribution is a rather weak effect. Unlike the degree of absolute truncation, the difference in outer radii between low- and high-mass host stars is maximised for large $\sigma_v$ (and large $n_\mathrm{c}$). Even in the extreme case of $n_\mathrm{c} = 5 \times 10^4$~pc$^{-3}$ and $\sigma_v= 5$~km/s the difference in $\bar{R}_\mathrm{out}$ between a brown dwarf and solar mass star is only $\sim 25\%$. A large sample of well-constrained PPD radii in such an environment would be required to detect any statistically significant differences between high- and low-mass stellar populations.  
\item Sub-structure can enhance encounter rates and therefore reduce disc radii within a stellar population in the short term. This is simply the statement that enhanced stellar densities result in increased truncation, and the canonical stellar density limit of $10^4$~pc$^{-3}$ should be seen as a threshold on the effective local stellar density (i.e. incorporating the role of sub-structure) in the context of observed environments.
\end{itemize}

The relevance of these conclusion depends on whether there exist any environments in which star-disc encounters are the dominant truncation mechanism within a cluster. We now use our results to compare to the photoevaporation rates we expect to find in the most dense cluster environments to establish the likeliness that this is ever the case.

\section{Photoevaporation}

\label{sec:photoevap}

In order to place our results regarding tidal truncation in context we need to compare with truncation induced by photoevaporation for a star which spends $3$~Myr in environments of a given $G_0$. Modelling disc evolution in this way requires knowledge of mass loss rates over a range of disc radii and $G_0$ values. The mass of the host star and disc also influences the photoevaporation rates, but this is addressed in detail by \citet{Haw18} and we therefore do not investigate here.  Instead we consider $0.1 \, M_\odot$ discs around solar mass stars and ask in what regimes are photoevaporation and tidal truncation dominant. Mass loss rates are greater for lower mass stars, therefore the loss rates quoted here represent a lower limit for the majority of a given stellar population.

\subsection{EUV vs. FUV Induced Mass-Loss}

\begin{figure}
	\includegraphics[width=\columnwidth]{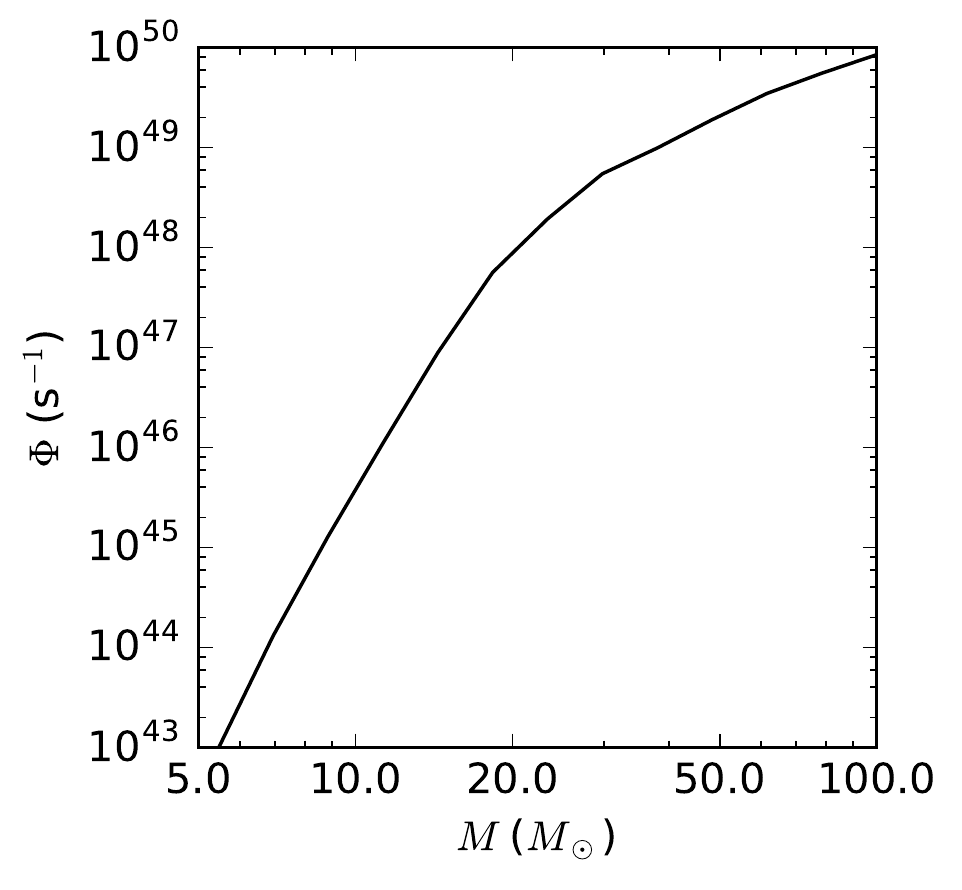}
    \caption{Number of EUV photons emitted from a star of a given mass. }
    \label{fig:EUVcts}
\end{figure}

\begin{figure}
	\includegraphics[width=\columnwidth]{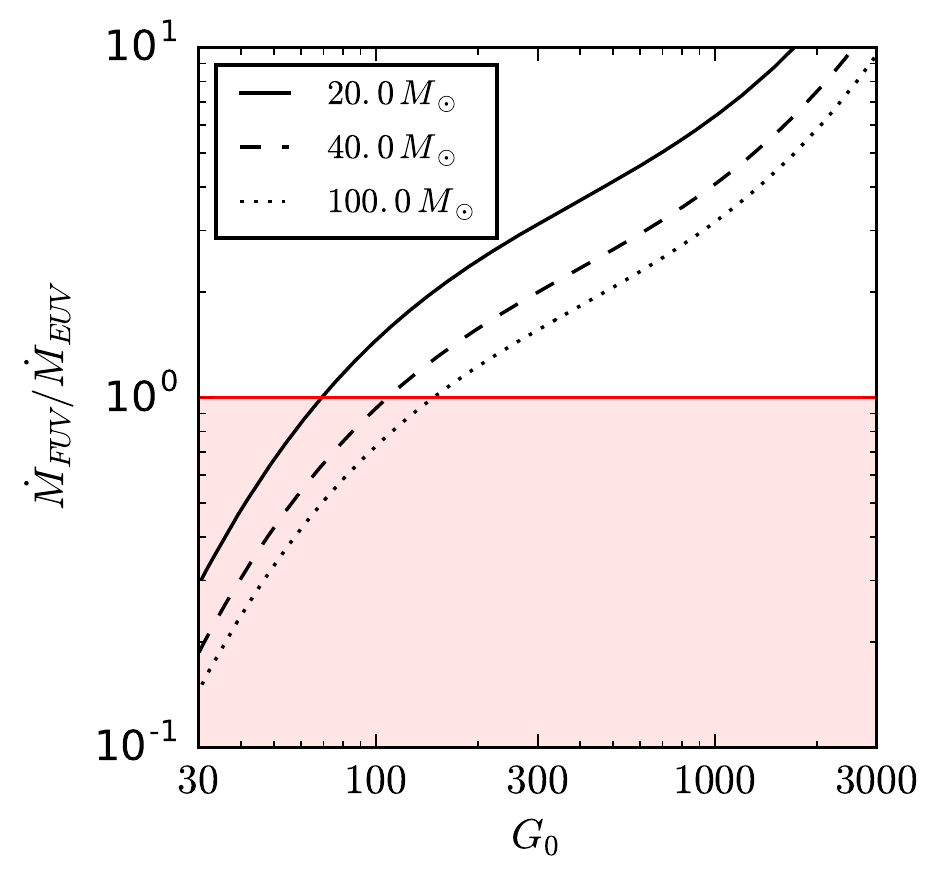}
    \caption{Ratio of the initial mass loss rates in a $0.1 \, M_\odot$ disc with $R_\mathrm{out} =100$~au around a $1 \, M_\odot$ induced by FUV versus EUV radiation. The region in which EUV photons induce greater mass loss ($\dot{M}_\mathrm{FUV}/\dot{M}_\mathrm{EUV}<1$) is shaded red. The cases for a $20\, M_\odot$, $40\, M_\odot$ and $100 \, M_\odot$ radiating source are shown. }
    \label{fig:EFrat}
\end{figure}

While we focus here on the mass loss rates induced by FUV photons, we also consider how our results might be altered by considering the ionising influence of the EUV photons. For outflows driven predominantly by the EUV, the FUV-induced photodissociation region (PDR) remains thin compared to the disc radius and the thermal pressure (and therefore the mass loss rate) is determined by photoionisation, rather than the heating of neutral gas as in the FUV case. 

FUV dominated flows can only occur when both the ionising EUV flux is relatively weak and there are sufficient FUV photons to heat gas to escape velocities. The escape velocity (and mass loss rate) is dependent on the temperature of the PDR, which is only weakly dependent on the FUV flux for $G_0  \sim 10^4-10^6$ (where the temperature of the PDR is $\sim 10^3$~K). Whether or not FUV can induce mass loss is dependent on $R_\mathrm{out}$ with respect to the critical radius 
$$
R_\mathrm{g}= \frac{G M_1}{c_s^2} \approx \frac{GM_1 \mu m_\mathrm{H}}{k_\mathrm{B}T} \approx \left( \frac{T}{1000 \, \mathrm{K}} \right)^{-1} \left( \frac{M_1}{M_\odot} \right) 140 \, \mathrm{au}
$$ for which the thermal energy is equal to the binding energy, where  $c_s(G_0)$ is the sound speed in the heated surface layer. In order for significant FUV induced photoevaporation to occur \citet{Ada04} find that $R_\mathrm{out}/R_\mathrm{g} \gtrsim 0.1$ is required. We apply the results of \citet{Fac16} to obtain expressions for the mass loss rate as a function of $R_\mathrm{out}$ for a range of $G_0$ values $30 < G_0 < 3000$.

 The above calculations assume that the effect of EUV radiation is restricted to radii in the flow that are outward of the sonic point in the FUV driven wind and that EUV radiation thus plays no part in setting the mass loss rate. If however the ionisation front lies close to the disc (i.e. the PDR region is thin) then the mass loss rate is set by conditions of ionisation balance close to the disc surface. In practice this means that the actual mass loss rate can be approximated by the maximum of the FUV rate and the mass loss rate resulting from EUV irradiation of an object equal in size to the disc (i.e. of size
$R_\mathrm{out}$). \citet{Joh98} find the expression for the EUV mass loss in the thin PDR limit is:
\begin{equation}
\label{eq:EUVloss}
\dot{M}_\mathrm{EUV} = 9.5 \times 10^{-9} \left(\frac{ f_r\Phi_{49}}{x_{15}^2} \right)^{1/2}  R_{12}^{3/2} \, M_\odot \, \mathrm{yr}^{-1}
\end{equation}
where 
$$
\Phi_{49} = \frac{ \Phi_i}{10^{49}\,  \mathrm{s}^{-1}}  \quad x_{15} = \frac{x_i}{10^{15} \, \mathrm{m}} \quad R_{12} =\frac{R_\mathrm{out}}{10^{12} \, \mathrm{m}}
$$ and $\Phi_i$ is the EUV photon luminosity of the source $i$ (shown as a function of stellar mass in Figure \ref{fig:EUVcts}), $x_i$ is the distance to the same source, and $f_r$ is the fraction of EUV photons which are not attenuated by the ISM, which we assume hereafter to be unity.

The ratio of the FUV to EUV loss rates are shown in Figure \ref{fig:EFrat}. For the range of $G_0$ values for which we have mass loss rates, the FUV dominates down to $G_0 \sim 100$. In reality there is a second region close to the star ($\lesssim 0.1$~pc, $G_0 \gtrsim 10^5$) at which the EUV is again expected to dominate, but our models do not cover this regime. This can be understood in that the EUV mass loss rate varies inversely with distance (Equation \ref{eq:EUVloss}) while the FUV mass loss rate plateaus at high $G_0 \gtrsim 10^4$ and then falls more steeply with declining  $G_0$ for $G_0 \lesssim 10^3$. These $G_0$ thresholds are lower than those found, for example, by \citet{Sto99} because the region where FUV is dominant is dependent on disc radius, and our $100$~au initial condition is larger than the disc radius considered in that study. While we will hereafter focus on the radius evolution of PPDs due to FUV radiation only, we note that this is effectively a lower limit on the rate of photoevaporation in the cluster environments we consider. This is sufficient for our purposes of comparing photoevaporation to tidal truncation in different regions.


\subsection{Disc Evolution}

To model the disc evolution we follow the method of \citet[][see also \citealt{And13}]{Cla07}, and model the disc subject to a combination of viscous evolution and photoevaporative mass loss from the outer edge. In brief, the viscous evolution is modelled with a parametrised viscosity that scales linearly with radius (corresponding to constant \citeauthor*{Sha73} $\alpha$ and a temperature scaling as $R^{-1/2}$) and the evolution is following on a one-dimensional grid equispaced in $R^{1/2}$. A zero torque inner boundary condition is applied; the cell that is deemed to be the instantaneous outer edge cell  is subject to both viscous outflow from the inwardly lying cell and a sink term for mass leaving in the wind. If the resultant of these leads to mass accumulation in the edge cell, the edge cell is advanced outwards. In the case that the edge cell is subject to net mass loss, we apply a threshold criterion at which we move the outer cell inwards, verifying that provided the threshold value is sufficiently low, the secular evolution is insensitive to its exact implementation.

We apply this disc evolution to a solar mass star, as this is the mass for which the largest datasets of PPD radii will be available, and we can compare to the corresponding calculations in Section \ref{sec:tidalavg}. To obtain a upper limit on the photoevaporation time-scale we choose the maximum initial disc mass that is compatible with gravitational stability ($0.1\, M_\odot$).  The viscosity is normalised so that the initial accretion rate onto the star is $7 \times 10^{-8} \, M_\odot$~yr$^{-1}$, consistent with the upper end of the accretion rate distribution for solar mass stars \citep{Man16}.

\subsection{PPD Destruction Time-scale}

\begin{figure}
	\includegraphics[width=\columnwidth]{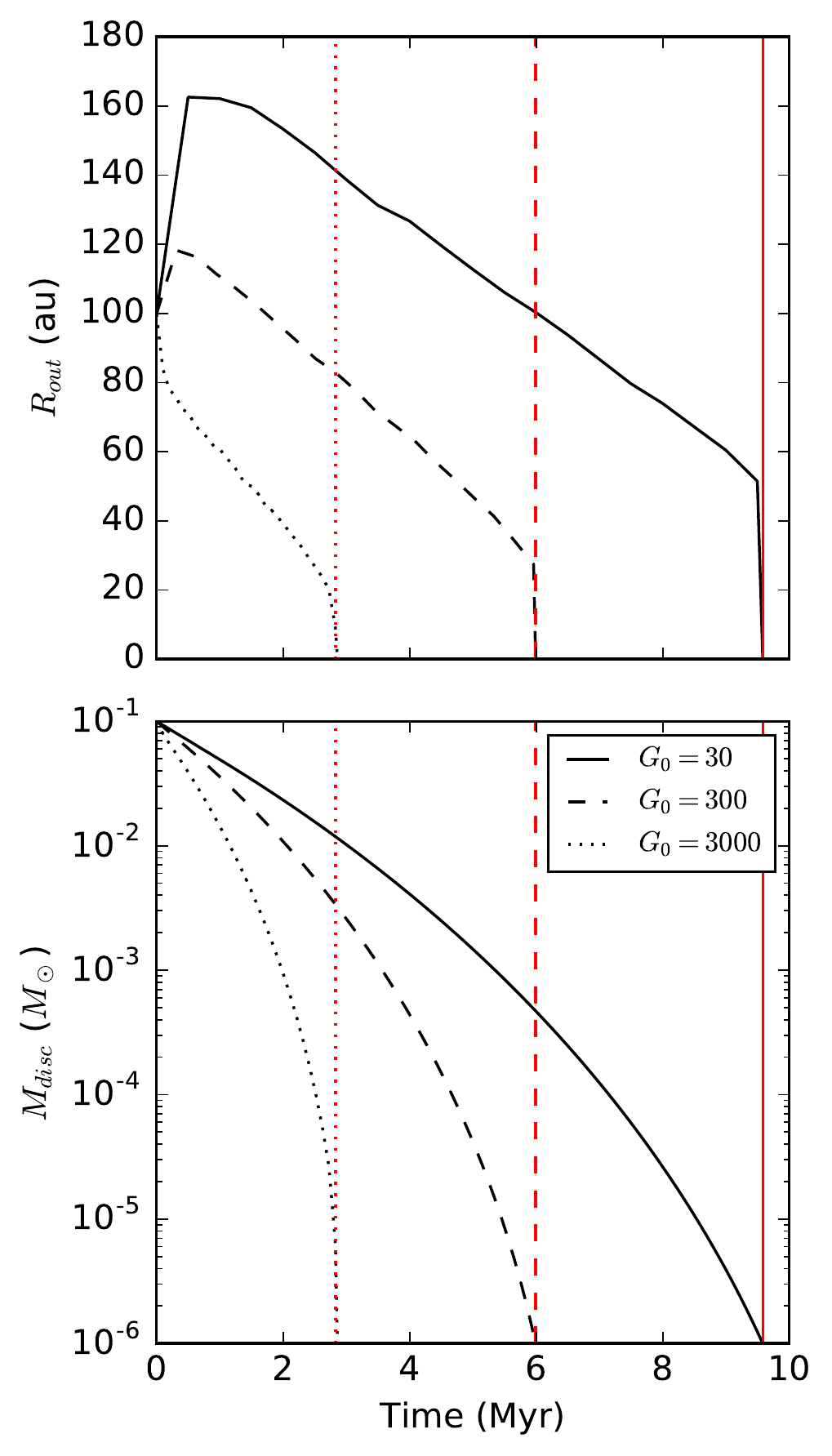}
    \caption{Outer radius (top) and mass (bottom) evolution of a $0.1 \, M_\odot$ PPD around a $1\, M_\odot$ star for $G_0= 30$ (solid), $300$ (dashed) and $3000$ (dotted). We have marked our definition of the photoevaporation induced disc destruction time-scale $\tau_\mathrm{phot.}(G_0)$ as a vertical red line in each case.  }
    \label{fig:g0revoldisc}
\end{figure}

\label{sec:phottime}
\begin{figure}
	\includegraphics[width=\columnwidth]{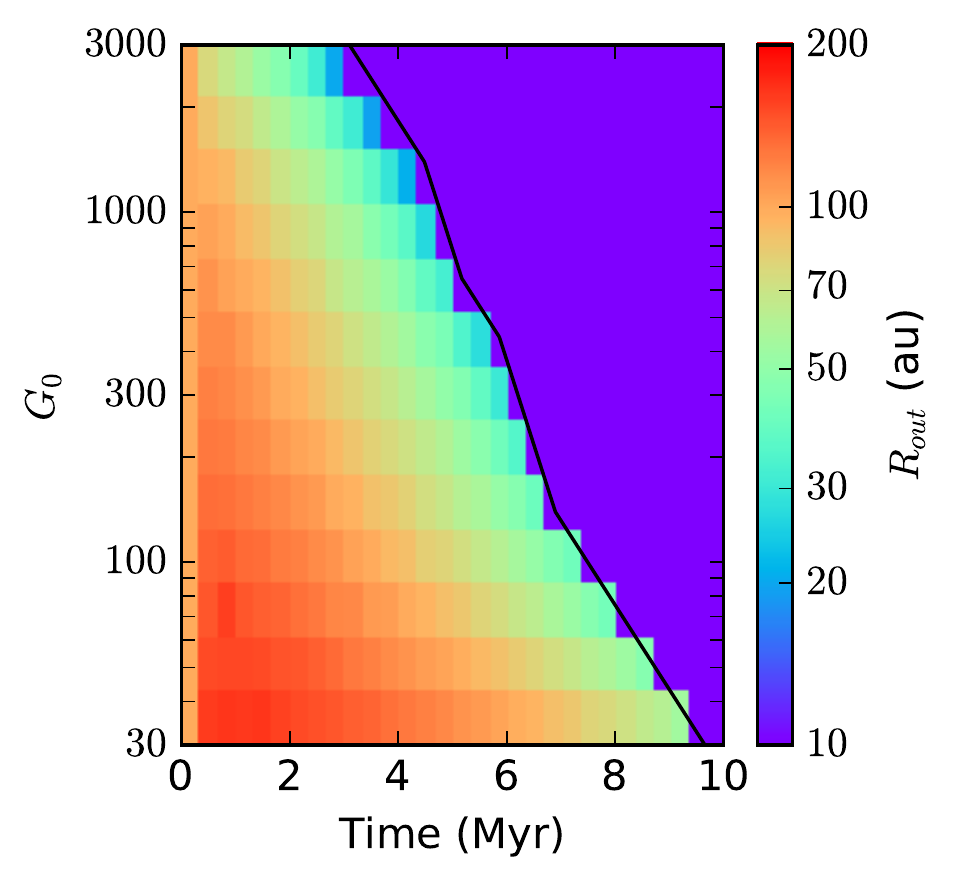}
    \caption{The evolution of the outer radius of a $0.1 \, M_\odot$ PPD around a $1\, M_\odot$ star in different (constant) FUV flux environments. The contour follows the time at which the disc is considered to be `destroyed', where $R_\mathrm{out}<10$~au or $M_\mathrm{disc}< 10^{-6}$~$M_\odot$.}
    \label{fig:g0rdisc}
\end{figure}

We apply our treatment of the disc radius evolution to a $0.1\, M_\odot$ disc around a $1\, M_\odot$ host star in Figure \ref{fig:g0revoldisc}. In the radius evolution we see that for a given initial disc profile, the disc shrinks  throughout its evolution for $G_0 = 3000$ but initially expands outwards for $G_0  \lesssim 300$ until the mass loss rate (which increases with $R_\mathrm{out}$) balances the viscous expansion. In this latter case the PPD then eventually shrinks once the disc has been significantly drained by both photoevaporation and accretion. The photoevaporation accelerates disc destruction even when the disc is very compact because it prevents the disc from viscously re-expanding by removing the material at the outer edge, thereby suppressing the evolution time-scale.

The time-scale $\tau_\mathrm{phot.}$ over which the FUV flux destroys a PPD is defined to be the time at which the disc is depleted such that $R_\mathrm{out} <10$~au or the mass falls below $10^{-6}\, M_\odot$ (in practice the disc lifetime is usually dictated by the latter). Such a definition is appropriate both because it represents a lower bound of the detectability of PPDs and because the disc does not persist long at low masses/radii as shown in Figure \ref{fig:g0revoldisc}. The value $\tau_\mathrm{phot.}$ is shown as a contour in Figure \ref{fig:g0rdisc} and we find that it varies between  $3-10$~Myr in the range $30-3000\, G_0$. 

Our definition of $\tau_\mathrm{phot.}$ is conservative (i.e. we define the time-scale over which photoevaporation occurs to be the time-scale for severe truncation). This is because our aim is to compare regions of dominance of the two truncation mechanisms (tidal encounters and photoevaporation) which act in different ways upon a disc population. Encounters are stochastic, and therefore cause a distribution of outer radii. In contrast, assuming all discs have the same initial conditions, the effect of FUV flux has a consistent effect on all such discs in the same environment. Therefore, in order to estimate the time-scale over which tidal encounters are irrelevant to the evolution of the PPD population as a whole, we choose $\tau_\mathrm{phot.}$ to be the period over which the disc is severely truncated (practically equivalent to the time-scale of complete truncation as shown in Figure \ref{fig:g0revoldisc}).

\section{Tidal Truncation vs. Photoevaporation}
\label{sec:comparison}

\begin{figure*}
	\includegraphics[width=\textwidth]{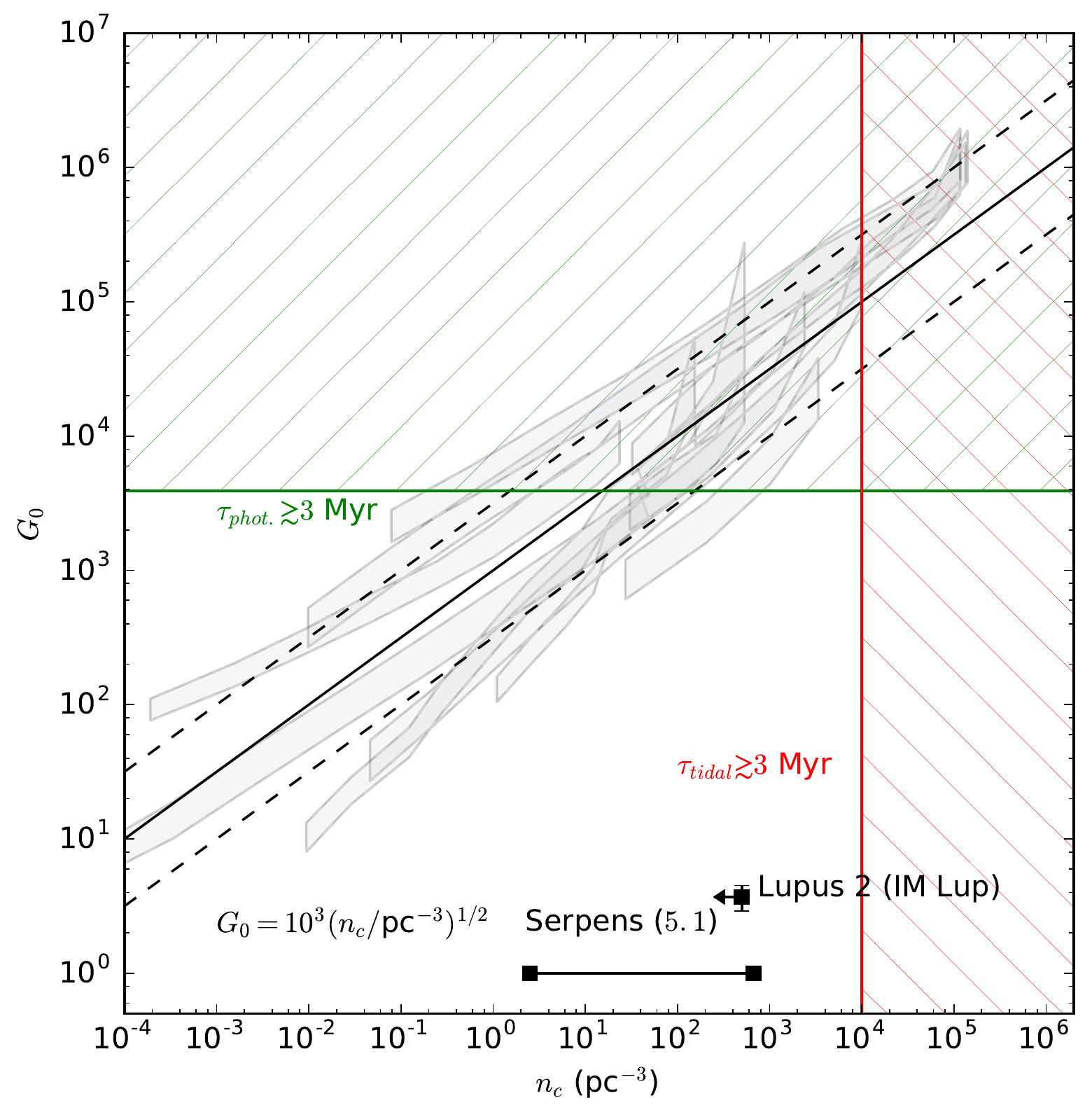}
    \caption{Cluster contours in $n_\mathrm{c}-G_0$ space as in Figure \ref{fig:nvsf} are shown here in grey. The horizontal green line shows the minimum $G_0$ such that the $0.1\, M_\odot$ disc around a $1 \, M_\odot$ star will be completely destroyed by photoevaporation within $3$~Myr. The vertical red line delineates the approximate regimes in which the number density is sufficient to produce significant tidal truncation for a $100$~au disc within $3$~Myr. The solid black line follows $G_0 = 10^3 (n_\mathrm{c}/\mathrm{pc}^{-3})^{1/2}$, with dashed lines showing $1$~dex around this value. The number in brackets next to the cluster name represents the assumed maximum mass in the cluster $m_\mathrm{max}$, which may be observed or predicted (see text for details).}
    \label{fig:nvsf_2}
\end{figure*}

In Figure \ref{fig:nvsf_2}  we have marked the flux limit for which we expect a $0.1\, M_\odot$ PPD around a $1\, M_\odot$ star to survive for $3$~Myr. We have further marked our approximate lower number density limit required to see significant tidal truncation in a disc population over the same period ($\sim 10^4$~pc$^{-3}$). As discussed in Section \ref{sec:tidalavg} this limit is moderately dependent on the local velocity dispersion, and some fraction of a population with $n_\mathrm{c} < 10^4$~pc$^{-3}$ will experience chance close encounters. 

We find that in all regions where significant tidal truncation occurs within $3$~Myr ($\tau_\mathrm{tidal} \lesssim 3$~Myr) also correspond to regions in which the FUV flux is sufficient to destroy the disc over this time-scale ($\tau_\mathrm{phot.} < 3$~Myr). In other words, we do not find any clusters or associations which contain environments occupying the bottom right of Figure \ref{fig:nvsf_2}. Therefore we conclude that there is no evidence supporting tidal truncation as a dominant mechanism influencing PPD evolution in real clusters. However, some caveats and possibilities are due discussion:

\begin{itemize}
\item Firstly, we acknowledge that our cluster sample is not complete. It is possible that there exist clusters with a low $m_\mathrm{max}$ and enhanced stellar densities in which tidal encounters are important for disc evolution. However, we do not find any local examples of such environments. Given that there is little variation in the flux-density profiles of the most massive clusters (Figure \ref{fig:nvsf_2}) we expect this can only be the case in clusters with $M_\mathrm{clust} \ll 10^3 \, M_\odot$. There is currently no evidence for such star forming environments in which encounters should dominate over photoevaporation as a disc truncation mechanism.

\item In present cluster environments past sub-structure might have enhanced number densities, and thus the encounter rate. However, based on the results in Figure \ref{fig:nvsf_2}, the degree of sub-structure must be such that local number densities are increased by more than two orders of magnitude to produce any regions in which tidal truncation is significant. Equally, an enhancement in number density will by definition reduce the distance between stars, and hence also increase the local FUV flux. Furthermore sub-structure is short-lived and corresponds to the highly embedded phase of star formation which is less well quantified than the (few Myr old) environments shown in Figures \ref{fig:nvsf} and \ref{fig:nvsf_2}. We emphasise that our calculations are designed to study the long term attrition of discs due to the influence of the mean environment on time-scales of Myr and that - in this case - dynamical interactions taking place within small dense multiple systems would be instead considered as providing disc initial conditions.

\item Extinction within young clusters can reduce the effective $G_0$ experienced by PPDs. Based on Figure \ref{fig:nvsf_2} the FUV flux would have to be reduced by a factor $\gtrsim 50$ to leave regions in a regime of tidally induced truncation. \citet{Car89} parametrise extinction as a function of wavelength, and in the FUV it is estimated at $A_\mathrm{FUV}/A_V \approx 2.7$. The column density of hydrogen, $N_H$, required for $1$~magnitude extinction is $N_H/A_V \approx 1.8 \times 10^{21}$~cm$^{-2}$~mag$^{-1}$ \citep{Pre95}. For a factor $50$ reduction in FUV flux, this corresponds to $\sim 10$~magnitudes, or $A_V \approx 3.6^m$ and $N_H \approx 6.5 \times 10^{21}$~cm$^{-2}$. If the gas distribution is uniform over $2$~pc this requires a volume density $n_H \sim 10^3$~cm$^{-3}$, which is high for a GMC \citep{Sol87}. Such extinction is observed, for example toward Cygnus OB2 where $A_V \sim 4^m-7^m$ is found, although this is likely due to foreground as well as internal gas \citep{Wri10, Gua12}. We further note that gas in stellar clusters tends to be clumpy, which reduces the efficiency with which it attenuates incident flux, and that the lifetime of the embedded phase is similar to the lifetime of a PPD \citep[$\sim 2-3$~Myr - see][]{Lada03}. Overall, while extinction may significantly reduce the effect of photoevaporation on a PPD population, it is not likely to be sufficient to make tidal encounters the dominant truncation mechanism in a local environment.
\item For the region in the top right of Figure \ref{fig:nvsf_2} (where both $\tau_\mathrm{phot.}, \, \tau_\mathrm{tidal} \lesssim 3$~Myr) we still do not expect tidal encounters to shape the outer radius distribution for three reasons. Firstly, the definition of $\tau_\mathrm{phot.}$ is such that the discs are completely destroyed by external photoevaporation, while $\tau_\mathrm{tidal}$ is the time-scale on which a PPD population might experience only mild tidally induced truncation. Secondly, the regions where both occur are spatially small, only existing in the very core $\sim 0.1$~pc of massive clusters and containing a small fraction of the overall stellar population. Thirdly, in these core regions with $G_0 \gtrsim 10^5$ the EUV will play the dominant role in mass loss, shortening the disc lifetime even for the extremely massive PPD ($0.1 \, M_\odot$) we have studied here.
\item Finally we note that our analysis of the effect of tidal truncation is based on the angle-averaged approach to individual encounters. In reality an additional scatter in outer radii of PPDs is expected (greater than that obtained using the \textit{Monte Carlo} approach discussed in Section \ref{sec:clust_numerics}). The fraction of discs which experience close encounters however, will remain unchanged. As our focus has been the effect of truncation mechanisms on a whole population of discs, we don't expect this scatter to alter our conclusions. Additionally, because we have chosen $\tau_\mathrm{phot.}$ as discussed in Section \ref{sec:phottime}, and the order of magnitude limit on $n_\mathrm{c}$ so that $\bar{R}_\mathrm{out}$ is only mildly truncated by encounters (see Figure \ref{fig:r_m_ratio}), the complete destruction of discs is expected to render the dispersion in encounter orientation irrelevant.

\end{itemize}

Ultimately we find that none of the environments we have selected are even particularly close to the region in which we would expect tidal encounters to play a significant role. Therefore, pending the discovery of radically different star formation environments, further investigation of the influence tidal encounters on the statistical properties of PPD populations is not likely to yield physically relevant insights. This does not preclude individual PPDs, or even small fractions of the PPD populations, from experiencing truncating encounters. However as a physical mechanism for disc truncation, photoevaporation is found to be far more efficient in real environments.

\section{Conclusions}
\label{sec:conclusion}

We have presented the first view of PPD truncation rates due to external photoevaporation and tidal encounters contextualised in terms of real local environments \citep[although for small clusters see][]{Ada06}. Examples of local clusters are modelled using observational density distributions and estimates for the maximum stellar mass. In this way we build a distribution of environments in terms of FUV flux ($G_0$) and stellar density ($n_\mathrm{c}$), and we seek to establish the dominant PPD truncation mechanism (if any) in each case. We note that our aim is not to reproduce the initial conditions for discs (which might for example be dependent on stellar multiplicity), but rather the long term evolution of a disc of a fixed initial radius in a local environment.

To this end we generate a full model for the angle-averaged tidal truncation radius of PPDs during a stellar encounter, and this model is fully summarised by Equations \ref{eq:fdef} through \ref{eq:fullmod}, with the parameters listed in Table \ref{tab:fittingparams}. For a given encounter closest approach distance $x_\mathrm{min}$, eccentricity $e_\mathrm{pert}$ and mass ratio $M_2/M_1$, the asymptotic truncation radius in the limit of the closest encounters is given by 
$$
R'_\mathrm{out} \approx 0.6 {e_\mathrm{pert}}^{0.11 f} \cdot  \left( \frac {M_2}{M_1} \right)^{-0.2} x_\mathrm{min}
$$ where $f=(M_2/M_1)^{-1/3}$. Applying this prescription statistically to local cluster environments, we find that stellar densities $n_\mathrm{c} \gtrsim 10^4$~pc$^{-3}$ are required to cause significant truncation of a PPD population within $3$~Myr. 

To compare this with photoevaporation time-scales we estimate the outer radius evolution of a viscously expanding PPD in environments of various $G_0$. We examine the case of a massive disc ($0.1\, M_\odot$) around a solar mass star with initial outer radius $\sim 100$~au and viscous accretion rate $7 \times 10^{-8}\,M_\odot$~yr$^{-1}$. We find that such a disc is destroyed within $3$~Myr for $G_0 \gtrsim 3000$. 

Having developed an understanding of the environments in which tidal truncation and external photoevaporation are significant, we examine real cluster environments and consider threshold values for both mechanisms. Our main findings from this comparison are as follows:

\begin{itemize}
\item In all the cluster environments discussed we find that regions for which the local number density is sufficient to yield significant truncation in a population of PPDs ($n_\mathrm{c} \gtrsim 10^4$~pc$^{-3}$) are also exposed to a strong FUV flux which causes complete destruction of even a massive disc within $3$~Myr ($G_0 \gtrsim 3000$). Therefore we conclude that environments in which tidal truncation shapes the distribution of outer radii are unlikely.

\item For massive clusters, the dispersion of FUV flux for a given local number density $n_\mathrm{c}$ is relatively small ($\lesssim 1$~dex). In particular, for $M_\mathrm{clust} \gtrsim  10^3 \, M_\odot$ we tentatively conclude that the FUV flux follows:
$$
G_0 = 1000 \left( \frac{n_c}{\mathrm{pc}^{-3}}\right)^{1/2}
$$
(Equation \ref{eq:nvg0}). Deviation from this relationship can occur due to very steep or shallow density profiles with radius within the cluster, or the presence of sub-structure. In general we find a greater fraction of stars in regions of stronger FUV flux than \citet{Ada06} because we consider more massive clusters than in that study.

\item In the less massive cluster regime $M_\mathrm{clust}\ll 10^3$, where the high-mass end of the IMF is not well sampled, the FUV flux is not defined by local number density. We find examples of low-mass clusters for which the FUV flux is much less than the number density would suggest according to Equation \ref{eq:nvg0}. However, even in these cases, the stellar densities are insufficient to induce significant tidal truncation of PPDs within $3$~Myr. Nonetheless, in principle low-mass environments with larger local stellar densities could exist.
\end{itemize}


In summary there is currently no evidence for star forming environments in which close encounters dictate disc extent, and it is likely that the total fraction of tidally truncated PPDs is small. Therefore star-disc interactions remain a secondary truncation mechanism. As greater samples of PPD properties are measured with ALMA, more complete disc radius and mass distributions as a function of distance from massive stars and local number density will become available to test these conclusions.

\section*{Acknowledgements}

We would like to thank the anonymous referee for a useful report which improved the clarity of this paper. We thank Marco Tazzari and Richard Booth for helpful discussion. AJW thanks  the  Science  and  Technology  Facilities  Council  (STFC)  for  their studentship. TJH is funded by an Imperial College Junior Research Fellowship. This work has been supported by the DISCSIM project, grant agreement 341137 funded by the European Research Council under ERC-2013-ADG. 




\bibliographystyle{mnras}
\bibliography{truncation} 

\begin{thebibliography}{}
\makeatletter
\relax
\def\mn@urlcharsother{\let\do\@makeother \do\$\do\&\do\#\do\^\do\_\do\%\do\~}
\def\mn@doi{\begingroup\mn@urlcharsother \@ifnextchar [ {\mn@doi@}
  {\mn@doi@[]}}
\def\mn@doi@[#1]#2{\def\@tempa{#1}\ifx\@tempa\@empty \href
  {http://dx.doi.org/#2} {doi:#2}\else \href {http://dx.doi.org/#2} {#1}\fi
  \endgroup}
\def\mn@eprint#1#2{\mn@eprint@#1:#2::\@nil}
\def\mn@eprint@arXiv#1{\href {http://arxiv.org/abs/#1} {{\tt arXiv:#1}}}
\def\mn@eprint@dblp#1{\href {http://dblp.uni-trier.de/rec/bibtex/#1.xml}
  {dblp:#1}}
\def\mn@eprint@#1:#2:#3:#4\@nil{\def\@tempa {#1}\def\@tempb {#2}\def\@tempc
  {#3}\ifx \@tempc \@empty \let \@tempc \@tempb \let \@tempb \@tempa \fi \ifx
  \@tempb \@empty \def\@tempb {arXiv}\fi \@ifundefined
  {mn@eprint@\@tempb}{\@tempb:\@tempc}{\expandafter \expandafter \csname
  mn@eprint@\@tempb\endcsname \expandafter{\@tempc}}}

\bibitem[\protect\citeauthoryear{{Adams}}{{Adams}}{2010}]{Ada10}
{Adams} F.~C.,  2010, \mn@doi [\araa] {10.1146/annurev-astro-081309-130830},
  48, 47

\bibitem[\protect\citeauthoryear{{Adams}, {Hollenbach}, {Laughlin}  \&
  {Gorti}}{{Adams} et~al.}{2004}]{Ada04}
{Adams} F.~C.,  {Hollenbach} D.,  {Laughlin} G.,   {Gorti} U.,  2004, \mn@doi
  [\apj] {10.1086/421989}, 611, 360

\bibitem[\protect\citeauthoryear{{Adams}, {Proszkow}, {Fatuzzo}  \&
  {Myers}}{{Adams} et~al.}{2006}]{Ada06}
{Adams} F.~C.,  {Proszkow} E.~M.,  {Fatuzzo} M.,   {Myers} P.~C.,  2006,
  \mn@doi [\apj] {10.1086/500393}, 641, 504

\bibitem[\protect\citeauthoryear{{Allison}, {Goodwin}, {Parker}, {de Grijs},
  {Portegies Zwart}  \& {Kouwenhoven}}{{Allison} et~al.}{2009}]{All09}
{Allison} R.~J.,  {Goodwin} S.~P.,  {Parker} R.~J.,  {de Grijs} R.,  {Portegies
  Zwart} S.~F.,   {Kouwenhoven} M.~B.~N.,  2009, \mn@doi [\apjl]
  {10.1088/0004-637X/700/2/L99}, 700, L99

\bibitem[\protect\citeauthoryear{{Alves de Oliveira} et~al.,}{{Alves de
  Oliveira} et~al.}{2013}]{Alv13}
{Alves de Oliveira} C.,  et~al., 2013, \mn@doi [\aap]
  {10.1051/0004-6361/201322402}, 559, A126

\bibitem[\protect\citeauthoryear{{Anderson}, {Adams}  \& {Calvet}}{{Anderson}
  et~al.}{2013}]{And13}
{Anderson} K.~R.,  {Adams} F.~C.,   {Calvet} N.,  2013, \mn@doi [\apj]
  {10.1088/0004-637X/774/1/9}, 774, 9

\bibitem[\protect\citeauthoryear{{Andrews} et~al.,}{{Andrews}
  et~al.}{2012}]{And12}
{Andrews} S.~M.,  et~al., 2012, \mn@doi [\apj] {10.1088/0004-637X/744/2/162},
  744, 162

\bibitem[\protect\citeauthoryear{{Ansdell} et~al.,}{{Ansdell}
  et~al.}{2016}]{Ans16}
{Ansdell} M.,  et~al., 2016, \mn@doi [\apj] {10.3847/0004-637X/828/1/46}, 828,
  46

\bibitem[\protect\citeauthoryear{{Ansdell}, {Williams}, {Manara}, {Miotello},
  {Facchini}, {van der Marel}, {Testi}  \& {van Dishoeck}}{{Ansdell}
  et~al.}{2017}]{Ans17}
{Ansdell} M.,  {Williams} J.~P.,  {Manara} C.~F.,  {Miotello} A.,  {Facchini}
  S.,  {van der Marel} N.,  {Testi} L.,   {van Dishoeck} E.~F.,  2017, \mn@doi
  [\aj] {10.3847/1538-3881/aa69c0}, 153, 240

\bibitem[\protect\citeauthoryear{{Ansdell} et~al.,}{{Ansdell}
  et~al.}{2018}]{Ans18}
{Ansdell} M.,  et~al., 2018, preprint (\mn@eprint {arXiv} {1803.05923})

\bibitem[\protect\citeauthoryear{{Anthony-Twarog}}{{Anthony-Twarog}}{1982}]{An%
t82}
{Anthony-Twarog} B.~J.,  1982, \mn@doi [\aj] {10.1086/113204}, 87, 1213

\bibitem[\protect\citeauthoryear{{Armitage}}{{Armitage}}{2000}]{Arm00}
{Armitage} P.~J.,  2000, \aap, 362, 968

\bibitem[\protect\citeauthoryear{{Barenfeld}, {Carpenter}, {Sargent}, {Isella}
  \& {Ricci}}{{Barenfeld} et~al.}{2017}]{Bar17}
{Barenfeld} S.~A.,  {Carpenter} J.~M.,  {Sargent} A.~I.,  {Isella} A.,
  {Ricci} L.,  2017, \mn@doi [\apj] {10.3847/1538-4357/aa989d}, 851, 85

\bibitem[\protect\citeauthoryear{{Bastian}}{{Bastian}}{2008}]{Bas08}
{Bastian} N.,  2008, \mn@doi [\mnras] {10.1111/j.1365-2966.2008.13775.x}, 390,
  759

\bibitem[\protect\citeauthoryear{{Bhandare}, {Breslau}  \&
  {Pfalzner}}{{Bhandare} et~al.}{2016}]{Bha16}
{Bhandare} A.,  {Breslau} A.,   {Pfalzner} S.,  2016, \mn@doi [\aap]
  {10.1051/0004-6361/201628086}, 594, A53

\bibitem[\protect\citeauthoryear{{Bik}, {Lenorzer}, {Kaper}, {Comer{\'o}n},
  {Waters}, {de Koter}  \& {Hanson}}{{Bik} et~al.}{2003}]{Bik03}
{Bik} A.,  {Lenorzer} A.,  {Kaper} L.,  {Comer{\'o}n} F.,  {Waters}
  L.~B.~F.~M.,  {de Koter} A.,   {Hanson} M.~M.,  2003, \mn@doi [\aap]
  {10.1051/0004-6361:20030301}, 404, 249

\bibitem[\protect\citeauthoryear{{Binney} \& {Tremaine}}{{Binney} \&
  {Tremaine}}{1987}]{Bin87}
{Binney} J.,  {Tremaine} S.,  1987, {Galactic dynamics}.
Princeton University Press

\bibitem[\protect\citeauthoryear{{Bonnell}, {Bate}  \& {Vine}}{{Bonnell}
  et~al.}{2003}]{Bon03}
{Bonnell} I.~A.,  {Bate} M.~R.,   {Vine} S.~G.,  2003, \mn@doi [\mnras]
  {10.1046/j.1365-8711.2003.06687.x}, 343, 413

\bibitem[\protect\citeauthoryear{{Breslau}, {Steinhausen}, {Vincke}  \&
  {Pfalzner}}{{Breslau} et~al.}{2014}]{Bre14}
{Breslau} A.,  {Steinhausen} M.,  {Vincke} K.,   {Pfalzner} S.,  2014, \mn@doi
  [\aap] {10.1051/0004-6361/201323043}, 565, A130

\bibitem[\protect\citeauthoryear{{Bressert} et~al.,}{{Bressert}
  et~al.}{2010}]{Bre10}
{Bressert} E.,  et~al., 2010, \mn@doi [\mnras]
  {10.1111/j.1745-3933.2010.00946.x}, 409, L54

\bibitem[\protect\citeauthoryear{{Caballero}}{{Caballero}}{2008a}]{Cab08}
{Caballero} J.~A.,  2008a, \mn@doi [\mnras] {10.1111/j.1365-2966.2007.12555.x},
  383, 375

\bibitem[\protect\citeauthoryear{{Caballero}}{{Caballero}}{2008b}]{Cab08a}
{Caballero} J.~A.,  2008b, \mn@doi [\aap] {10.1051/0004-6361:20077885}, 478,
  667

\bibitem[\protect\citeauthoryear{{Cardelli}, {Clayton}  \& {Mathis}}{{Cardelli}
  et~al.}{1989}]{Car89}
{Cardelli} J.~A.,  {Clayton} G.~C.,   {Mathis} J.~S.,  1989, \mn@doi [\apj]
  {10.1086/167900}, 345, 245

\bibitem[\protect\citeauthoryear{{Castelli} \& {Kurucz}}{{Castelli} \&
  {Kurucz}}{2004}]{Cas04}
{Castelli} F.,  {Kurucz} R.~L.,  2004, ArXiv Astrophysics e-prints

\bibitem[\protect\citeauthoryear{{Chambers}}{{Chambers}}{1999}]{Cha99}
{Chambers} J.~E.,  1999, \mn@doi [\mnras] {10.1046/j.1365-8711.1999.02379.x},
  304, 793

\bibitem[\protect\citeauthoryear{{Clark}, {Negueruela}, {Crowther}  \&
  {Goodwin}}{{Clark} et~al.}{2005}]{Cla05}
{Clark} J.~S.,  {Negueruela} I.,  {Crowther} P.~A.,   {Goodwin} S.~P.,  2005,
  \mn@doi [\aap] {10.1051/0004-6361:20042413}, 434, 949

\bibitem[\protect\citeauthoryear{{Clarke}}{{Clarke}}{2007}]{Cla07}
{Clarke} C.~J.,  2007, \mn@doi [\mnras] {10.1111/j.1365-2966.2007.11547.x},
  376, 1350

\bibitem[\protect\citeauthoryear{{Clarke} \& {Pringle}}{{Clarke} \&
  {Pringle}}{1993}]{Cla93}
{Clarke} C.~J.,  {Pringle} J.~E.,  1993, \mn@doi [\mnras]
  {10.1093/mnras/261.1.190}, 261, 190

\bibitem[\protect\citeauthoryear{{Clarkson}, {Ghez}, {Morris}, {Lu}, {Stolte},
  {McCrady}, {Do}  \& {Yelda}}{{Clarkson} et~al.}{2012}]{Cla12}
{Clarkson} W.~I.,  {Ghez} A.~M.,  {Morris} M.~R.,  {Lu} J.~R.,  {Stolte} A.,
  {McCrady} N.,  {Do} T.,   {Yelda} S.,  2012, \mn@doi [\apj]
  {10.1088/0004-637X/751/2/132}, 751, 132

\bibitem[\protect\citeauthoryear{{Cleeves}, {{\"O}berg}, {Wilner}, {Huang},
  {Loomis}, {Andrews}  \& {Czekala}}{{Cleeves} et~al.}{2016}]{Cle16}
{Cleeves} L.~I.,  {{\"O}berg} K.~I.,  {Wilner} D.~J.,  {Huang} J.,  {Loomis}
  R.~A.,  {Andrews} S.~M.,   {Czekala} I.,  2016, \mn@doi [\apj]
  {10.3847/0004-637X/832/2/110}, 832, 110

\bibitem[\protect\citeauthoryear{{Comer{\'o}n}}{{Comer{\'o}n}}{2008}]{Com08}
{Comer{\'o}n} F.,  2008, {The Lupus Clouds}.
p.~295

\bibitem[\protect\citeauthoryear{{Cox} et~al.,}{{Cox} et~al.}{2017}]{Cox17}
{Cox} E.~G.,  et~al., 2017, \mn@doi [\apj] {10.3847/1538-4357/aa97e2}, 851, 83

\bibitem[\protect\citeauthoryear{{Craig} \& {Krumholz}}{{Craig} \&
  {Krumholz}}{2013}]{Cra13}
{Craig} J.,  {Krumholz} M.~R.,  2013, \mn@doi [\apj]
  {10.1088/0004-637X/769/2/150}, 769, 150

\bibitem[\protect\citeauthoryear{{Dolan} \& {Mathieu}}{{Dolan} \&
  {Mathieu}}{2001}]{Dol01}
{Dolan} C.~J.,  {Mathieu} R.~D.,  2001, \mn@doi [\aj] {10.1086/319946}, 121,
  2124

\bibitem[\protect\citeauthoryear{{Dukes} \& {Krumholz}}{{Dukes} \&
  {Krumholz}}{2012}]{Duk12}
{Dukes} D.,  {Krumholz} M.~R.,  2012, \mn@doi [\apj]
  {10.1088/0004-637X/754/1/56}, 754, 56

\bibitem[\protect\citeauthoryear{{Dutrey}, {Guilloteau}, {Prato}, {Simon},
  {Duvert}, {Schuster}  \& {Menard}}{{Dutrey} et~al.}{1998}]{Dut98}
{Dutrey} A.,  {Guilloteau} S.,  {Prato} L.,  {Simon} M.,  {Duvert} G.,
  {Schuster} K.,   {Menard} F.,  1998, \aap, 338, L63

\bibitem[\protect\citeauthoryear{{Elson}, {Fall}  \& {Freeman}}{{Elson}
  et~al.}{1987}]{Els87}
{Elson} R.~A.~W.,  {Fall} S.~M.,   {Freeman} K.~C.,  1987, \mn@doi [\apj]
  {10.1086/165807}, 323, 54

\bibitem[\protect\citeauthoryear{{Erickson}, {Wilking}, {Meyer}, {Kim},
  {Sherry}  \& {Freeman}}{{Erickson} et~al.}{2015}]{Eri15}
{Erickson} K.~L.,  {Wilking} B.~A.,  {Meyer} M.~R.,  {Kim} J.~S.,  {Sherry} W.,
    {Freeman} M.,  2015, \mn@doi [\aj] {10.1088/0004-6256/149/3/103}, 149, 103

\bibitem[\protect\citeauthoryear{{Facchini}, {Clarke}  \& {Bisbas}}{{Facchini}
  et~al.}{2016}]{Fac16}
{Facchini} S.,  {Clarke} C.~J.,   {Bisbas} T.~G.,  2016, \mn@doi [\mnras]
  {10.1093/mnras/stw240}, 457, 3593

\bibitem[\protect\citeauthoryear{{Facchini}, {Birnstiel}, {Bruderer}  \& {van
  Dishoeck}}{{Facchini} et~al.}{2017}]{Fac17}
{Facchini} S.,  {Birnstiel} T.,  {Bruderer} S.,   {van Dishoeck} E.~F.,  2017,
  \mn@doi [\aap] {10.1051/0004-6361/201630329}, 605, A16

\bibitem[\protect\citeauthoryear{{Fatuzzo} \& {Adams}}{{Fatuzzo} \&
  {Adams}}{2008}]{Fat08}
{Fatuzzo} M.,  {Adams} F.~C.,  2008, \mn@doi [\apj] {10.1086/527469}, 675, 1361

\bibitem[\protect\citeauthoryear{{Foreman-Mackey}, {Hogg}, {Lang}  \&
  {Goodman}}{{Foreman-Mackey} et~al.}{2013}]{For13}
{Foreman-Mackey} D.,  {Hogg} D.~W.,  {Lang} D.,   {Goodman} J.,  2013, \mn@doi
  [PASP] {10.1086/670067}, 125, 306

\bibitem[\protect\citeauthoryear{{Getman}, {Feigelson}  \& {Kuhn}}{{Getman}
  et~al.}{2014}]{Get14}
{Getman} K.~V.,  {Feigelson} E.~D.,   {Kuhn} M.~A.,  2014, \mn@doi [\apj]
  {10.1088/0004-637X/787/2/109}, 787, 109

\bibitem[\protect\citeauthoryear{{Gieles}, {Larsen}, {Scheepmaker}, {Bastian},
  {Haas}  \& {Lamers}}{{Gieles} et~al.}{2006}]{Gie06}
{Gieles} M.,  {Larsen} S.~S.,  {Scheepmaker} R.~A.,  {Bastian} N.,  {Haas}
  M.~R.,   {Lamers} H.~J.~G.~L.~M.,  2006, \mn@doi [\aap]
  {10.1051/0004-6361:200500224}, 446, L9

\bibitem[\protect\citeauthoryear{{Gieles}, {Sana}  \& {Portegies
  Zwart}}{{Gieles} et~al.}{2010}]{Gie10}
{Gieles} M.,  {Sana} H.,   {Portegies Zwart} S.~F.,  2010, \mn@doi [\mnras]
  {10.1111/j.1365-2966.2009.15993.x}, 402, 1750

\bibitem[\protect\citeauthoryear{{Guarcello}, {Wright}, {Drake},
  {Garc{\'{\i}}a-Alvarez}, {Drew}, {Aldcroft}  \& {Kashyap}}{{Guarcello}
  et~al.}{2012}]{Gua12}
{Guarcello} M.~G.,  {Wright} N.~J.,  {Drake} J.~J.,  {Garc{\'{\i}}a-Alvarez}
  D.,  {Drew} J.~E.,  {Aldcroft} T.,   {Kashyap} V.~L.,  2012, \mn@doi [\apjs]
  {10.1088/0067-0049/202/2/19}, 202, 19

\bibitem[\protect\citeauthoryear{{Guarcello} et~al.,}{{Guarcello}
  et~al.}{2016}]{Gua16}
{Guarcello} M.~G.,  et~al., 2016, preprint (\mn@eprint {arXiv} {1605.01773})

\bibitem[\protect\citeauthoryear{{Guilloteau} \& {Dutrey}}{{Guilloteau} \&
  {Dutrey}}{1998}]{Gui98}
{Guilloteau} S.,  {Dutrey} A.,  1998, \aap, 339, 467

\bibitem[\protect\citeauthoryear{{Haisch}, {Lada}  \& {Lada}}{{Haisch}
  et~al.}{2000}]{Hai00}
{Haisch} Jr. K.~E.,  {Lada} E.~A.,   {Lada} C.~J.,  2000, \mn@doi [\aj]
  {10.1086/301521}, 120, 1396

\bibitem[\protect\citeauthoryear{{Haisch}, {Lada}, {Pi{\~n}a}, {Telesco}  \&
  {Lada}}{{Haisch} et~al.}{2001}]{Hai01}
{Haisch} Jr. K.~E.,  {Lada} E.~A.,  {Pi{\~n}a} R.~K.,  {Telesco} C.~M.,
  {Lada} C.~J.,  2001, \mn@doi [\aj] {10.1086/319397}, 121, 1512

\bibitem[\protect\citeauthoryear{{Hall}, {Clarke}  \& {Pringle}}{{Hall}
  et~al.}{1996}]{Hal96}
{Hall} S.~M.,  {Clarke} C.~J.,   {Pringle} J.~E.,  1996, \mn@doi [\mnras]
  {10.1093/mnras/278.2.303}, 278, 303

\bibitem[\protect\citeauthoryear{{Harvey}, {Mer{\'{\i}}n}, {Huard}, {Rebull},
  {Chapman}, {Evans}  \& {Myers}}{{Harvey} et~al.}{2007}]{Har07}
{Harvey} P.,  {Mer{\'{\i}}n} B.,  {Huard} T.~L.,  {Rebull} L.~M.,  {Chapman}
  N.,  {Evans} II N.~J.,   {Myers} P.~C.,  2007, \mn@doi [\apj]
  {10.1086/518646}, 663, 1149

\bibitem[\protect\citeauthoryear{{Haworth}, {Facchini}, {Clarke}  \&
  {Cleeves}}{{Haworth} et~al.}{2017}]{Haw17b}
{Haworth} T.~J.,  {Facchini} S.,  {Clarke} C.~J.,   {Cleeves} L.~I.,  2017,
  \mn@doi [\mnras] {10.1093/mnrasl/slx037}, 468, L108

\bibitem[\protect\citeauthoryear{{Haworth}, {Facchini}, {Clarke}  \&
  {Mohanty}}{{Haworth} et~al.}{2018}]{Haw18}
{Haworth} T.~J.,  {Facchini} S.,  {Clarke} C.~J.,   {Mohanty} S.,  2018,
  \mn@doi [\mnras] {10.1093/mnras/sty168}

\bibitem[\protect\citeauthoryear{{Hern{\'a}ndez}, {Morales-Calderon}, {Calvet},
  {Hartmann}, {Muzerolle}, {Gutermuth}, {Luhman}  \&
  {Stauffer}}{{Hern{\'a}ndez} et~al.}{2010}]{Her10}
{Hern{\'a}ndez} J.,  {Morales-Calderon} M.,  {Calvet} N.,  {Hartmann} L.,
  {Muzerolle} J.,  {Gutermuth} R.,  {Luhman} K.~L.,   {Stauffer} J.,  2010,
  \mn@doi [\apj] {10.1088/0004-637X/722/2/1226}, 722, 1226

\bibitem[\protect\citeauthoryear{{Hillenbrand} \& {Hartmann}}{{Hillenbrand} \&
  {Hartmann}}{1998}]{Hil98}
{Hillenbrand} L.~A.,  {Hartmann} L.~W.,  1998, \mn@doi [\apj] {10.1086/305076},
  492, 540

\bibitem[\protect\citeauthoryear{{Johnstone}, {Fabian}  \&
  {Taylor}}{{Johnstone} et~al.}{1998}]{Joh98}
{Johnstone} R.~M.,  {Fabian} A.~C.,   {Taylor} G.~B.,  1998, \mn@doi [\mnras]
  {10.1046/j.1365-8711.1998.01675.x}, 298, 854

\bibitem[\protect\citeauthoryear{{Kim}, {Clarke}, {Fang}  \& {Facchini}}{{Kim}
  et~al.}{2016}]{Kim16}
{Kim} J.~S.,  {Clarke} C.~J.,  {Fang} M.,   {Facchini} S.,  2016, \mn@doi
  [\apjl] {10.3847/2041-8205/826/1/L15}, 826, L15

\bibitem[\protect\citeauthoryear{{Kraus} et~al.,}{{Kraus} et~al.}{2009}]{Kra09}
{Kraus} S.,  et~al., 2009, \mn@doi [\aap] {10.1051/0004-6361/200810368}, 497,
  195

\bibitem[\protect\citeauthoryear{{Kroupa}, {Tout}  \& {Gilmore}}{{Kroupa}
  et~al.}{1993}]{Kro93}
{Kroupa} P.,  {Tout} C.~A.,   {Gilmore} G.,  1993, \mn@doi [\mnras]
  {10.1093/mnras/262.3.545}, 262, 545

\bibitem[\protect\citeauthoryear{{Kruijssen}}{{Kruijssen}}{2012}]{Kru12}
{Kruijssen} J.~M.~D.,  2012, \mn@doi [\mnras]
  {10.1111/j.1365-2966.2012.21923.x}, 426, 3008

\bibitem[\protect\citeauthoryear{{Kruijssen}, {Maschberger}, {Moeckel},
  {Clarke}, {Bastian}  \& {Bonnell}}{{Kruijssen} et~al.}{2012}]{Kru12b}
{Kruijssen} J.~M.~D.,  {Maschberger} T.,  {Moeckel} N.,  {Clarke} C.~J.,
  {Bastian} N.,   {Bonnell} I.~A.,  2012, \mn@doi [\mnras]
  {10.1111/j.1365-2966.2011.19748.x}, 419, 841

\bibitem[\protect\citeauthoryear{{Kutner}, {Evans}  \& {Tucker}}{{Kutner}
  et~al.}{1976}]{Kut76}
{Kutner} M.~L.,  {Evans} II N.~J.,   {Tucker} K.~D.,  1976, \mn@doi [\apj]
  {10.1086/154739}, 209, 452

\bibitem[\protect\citeauthoryear{{Lada} \& {Kylafis}}{{Lada} \&
  {Kylafis}}{1999}]{Lad99}
{Lada} C.~J.,  {Kylafis} N.~D.,  eds, 1999, {The Origin of Stars and Planetary
  Systems}  NATO Advanced Science Institutes (ASI) Series C Vol. 540

\bibitem[\protect\citeauthoryear{{Lada} \& {Lada}}{{Lada} \&
  {Lada}}{2003}]{Lada03}
{Lada} C.~J.,  {Lada} E.~A.,  2003, \mn@doi [\araa]
  {10.1146/annurev.astro.41.011802.094844}, 41, 57

\bibitem[\protect\citeauthoryear{{Law}, {Ricci}, {Andrews}, {Wilner}  \&
  {Qi}}{{Law} et~al.}{2017}]{Law17}
{Law} C.~J.,  {Ricci} L.,  {Andrews} S.~M.,  {Wilner} D.~J.,   {Qi} C.,  2017,
  \mn@doi [\aj] {10.3847/1538-3881/aa9752}, 154, 255

\bibitem[\protect\citeauthoryear{{Lee}, {Seon}  \& {Jo}}{{Lee}
  et~al.}{2015}]{Lee15}
{Lee} D.,  {Seon} K.-I.,   {Jo} Y.-S.,  2015, \mn@doi [\apj]
  {10.1088/0004-637X/806/2/274}, 806, 274

\bibitem[\protect\citeauthoryear{{Lenorzer}, {Bik}, {de Koter}, {Kurtz},
  {Waters}, {Kaper}, {Jones}  \& {Geballe}}{{Lenorzer} et~al.}{2004}]{Len04}
{Lenorzer} A.,  {Bik} A.,  {de Koter} A.,  {Kurtz} S.~E.,  {Waters}
  L.~B.~F.~M.,  {Kaper} L.,  {Jones} C.~E.,   {Geballe} T.~R.,  2004, \mn@doi
  [\aap] {10.1051/0004-6361:20031580}, 414, 245

\bibitem[\protect\citeauthoryear{{Levine}, {Steinhauer}, {Elston}  \&
  {Lada}}{{Levine} et~al.}{2006}]{Lev06}
{Levine} J.~L.,  {Steinhauer} A.,  {Elston} R.~J.,   {Lada} E.~A.,  2006,
  \mn@doi [\apj] {10.1086/504964}, 646, 1215

\bibitem[\protect\citeauthoryear{{Maddalena} \& {Morris}}{{Maddalena} \&
  {Morris}}{1987}]{Mad87}
{Maddalena} R.~J.,  {Morris} M.,  1987, \mn@doi [\apj] {10.1086/165818}, 323,
  179

\bibitem[\protect\citeauthoryear{{Manara} et~al.,}{{Manara}
  et~al.}{2016}]{Man16}
{Manara} C.~F.,  et~al., 2016, \mn@doi [\aap] {10.1051/0004-6361/201628549},
  591, L3

\bibitem[\protect\citeauthoryear{{Mann}, {Andrews}, {Eisner}, {Williams},
  {Meyer}, {Di Francesco}, {Carpenter}  \& {Johnstone}}{{Mann}
  et~al.}{2015}]{Man15}
{Mann} R.~K.,  {Andrews} S.~M.,  {Eisner} J.~A.,  {Williams} J.~P.,  {Meyer}
  M.~R.,  {Di Francesco} J.,  {Carpenter} J.~M.,   {Johnstone} D.,  2015,
  \mn@doi [\apj] {10.1088/0004-637X/802/2/77}, 802, 77

\bibitem[\protect\citeauthoryear{{Maschberger} \& {Clarke}}{{Maschberger} \&
  {Clarke}}{2008}]{Mas08}
{Maschberger} T.,  {Clarke} C.~J.,  2008, \mn@doi [\mnras]
  {10.1111/j.1365-2966.2008.13903.x}, 391, 711

\bibitem[\protect\citeauthoryear{{Mauc{\'o}} et~al.,}{{Mauc{\'o}}
  et~al.}{2016}]{Mau16}
{Mauc{\'o}} K.,  et~al., 2016, \mn@doi [\apj] {10.3847/0004-637X/829/1/38},
  829, 38

\bibitem[\protect\citeauthoryear{{Mayne} \& {Naylor}}{{Mayne} \&
  {Naylor}}{2008}]{May08}
{Mayne} N.~J.,  {Naylor} T.,  2008, \mn@doi [\mnras]
  {10.1111/j.1365-2966.2008.13025.x}, 386, 261

\bibitem[\protect\citeauthoryear{{Mengel} \& {Tacconi-Garman}}{{Mengel} \&
  {Tacconi-Garman}}{2007}]{Men07}
{Mengel} S.,  {Tacconi-Garman} L.~E.,  2007, \mn@doi [\aap]
  {10.1051/0004-6361:20066717}, 466, 151

\bibitem[\protect\citeauthoryear{{Mer{\'{\i}}n} et~al.,}{{Mer{\'{\i}}n}
  et~al.}{2008}]{Mer08}
{Mer{\'{\i}}n} B.,  et~al., 2008, \mn@doi [\apjs] {10.1086/588042}, 177, 551

\bibitem[\protect\citeauthoryear{{Meyer}}{{Meyer}}{1996}]{Mey96}
{Meyer} M.~R.,  1996, PhD thesis, Max-Planck-Institut f{\"u}r Astronomie,
  K{\"o}nigstuhl 17, D-69117 Heidelberg, Germany

\bibitem[\protect\citeauthoryear{{Mu{\~n}oz}, {Kratter}, {Vogelsberger},
  {Hernquist}  \& {Springel}}{{Mu{\~n}oz} et~al.}{2015}]{Mun15}
{Mu{\~n}oz} D.~J.,  {Kratter} K.,  {Vogelsberger} M.,  {Hernquist} L.,
  {Springel} V.,  2015, \mn@doi [\mnras] {10.1093/mnras/stu2220}, 446, 2010

\bibitem[\protect\citeauthoryear{{Murray}}{{Murray}}{2011}]{Mur11}
{Murray} N.,  2011, \mn@doi [\apj] {10.1088/0004-637X/729/2/133}, 729, 133

\bibitem[\protect\citeauthoryear{{Nakajima}, {Tamura}, {Oasa}  \&
  {Nakajima}}{{Nakajima} et~al.}{2000}]{Nak00}
{Nakajima} Y.,  {Tamura} M.,  {Oasa} Y.,   {Nakajima} T.,  2000, \mn@doi [\aj]
  {10.1086/301222}, 119, 873

\bibitem[\protect\citeauthoryear{{Olczak}, {Pfalzner}  \& {Spurzem}}{{Olczak}
  et~al.}{2006}]{Olc06}
{Olczak} C.,  {Pfalzner} S.,   {Spurzem} R.,  2006, \mn@doi [\apj]
  {10.1086/501044}, 642, 1140

\bibitem[\protect\citeauthoryear{{Olczak}, {Pfalzner}  \& {Eckart}}{{Olczak}
  et~al.}{2010}]{Olc10}
{Olczak} C.,  {Pfalzner} S.,   {Eckart} A.,  2010, \mn@doi [\aap]
  {10.1051/0004-6361/200912641}, 509, A63

\bibitem[\protect\citeauthoryear{{Ostriker}}{{Ostriker}}{1994}]{Ost94}
{Ostriker} E.~C.,  1994, \mn@doi [\apj] {10.1086/173890}, 424, 292

\bibitem[\protect\citeauthoryear{{Parker}, {Goodwin}  \& {Allison}}{{Parker}
  et~al.}{2011}]{Par11}
{Parker} R.~J.,  {Goodwin} S.~P.,   {Allison} R.~J.,  2011, \mn@doi [\mnras]
  {10.1111/j.1365-2966.2011.19646.x}, 418, 2565

\bibitem[\protect\citeauthoryear{{Peterson} \& {Megeath}}{{Peterson} \&
  {Megeath}}{2008}]{Pet08}
{Peterson} D.~E.,  {Megeath} S.~T.,  2008, {The Orion Molecular Cloud 2/3 and
  NGC 1977 Regions}.
p.~590

\bibitem[\protect\citeauthoryear{{Pfalzner} \& {Kaczmarek}}{{Pfalzner} \&
  {Kaczmarek}}{2013}]{Pfa13}
{Pfalzner} S.,  {Kaczmarek} T.,  2013, \mn@doi [\aap]
  {10.1051/0004-6361/201322134}, 559, A38

\bibitem[\protect\citeauthoryear{{Pfalzner}, {Vogel}, {Scharw{\"a}chter}  \&
  {Olczak}}{{Pfalzner} et~al.}{2005}]{Pfa05}
{Pfalzner} S.,  {Vogel} P.,  {Scharw{\"a}chter} J.,   {Olczak} C.,  2005,
  \mn@doi [\aap] {10.1051/0004-6361:20042467}, 437, 967

\bibitem[\protect\citeauthoryear{{Pfalzner}, {Olczak}  \& {Eckart}}{{Pfalzner}
  et~al.}{2006}]{Pfa06}
{Pfalzner} S.,  {Olczak} C.,   {Eckart} A.,  2006, \mn@doi [\aap]
  {10.1051/0004-6361:20064905}, 454, 811

\bibitem[\protect\citeauthoryear{{Pi{\'e}tu}, {Guilloteau}, {Di Folco},
  {Dutrey}  \& {Boehler}}{{Pi{\'e}tu} et~al.}{2014}]{Pie14}
{Pi{\'e}tu} V.,  {Guilloteau} S.,  {Di Folco} E.,  {Dutrey} A.,   {Boehler} Y.,
   2014, \mn@doi [\aap] {10.1051/0004-6361/201322388}, 564, A95

\bibitem[\protect\citeauthoryear{{Portegies Zwart}, {McMillan}  \&
  {Gieles}}{{Portegies Zwart} et~al.}{2010}]{Zwa10}
{Portegies Zwart} S.~F.,  {McMillan} S.~L.~W.,   {Gieles} M.,  2010, \mn@doi
  [\araa] {10.1146/annurev-astro-081309-130834}, 48, 431

\bibitem[\protect\citeauthoryear{{Predehl} \& {Schmitt}}{{Predehl} \&
  {Schmitt}}{1995}]{Pre95}
{Predehl} P.,  {Schmitt} J.~H.~M.~M.,  1995, \aap, 293, 889

\bibitem[\protect\citeauthoryear{{Preibisch}, {Brown}, {Bridges}, {Guenther}
  \& {Zinnecker}}{{Preibisch} et~al.}{2002}]{Pre02}
{Preibisch} T.,  {Brown} A.~G.~A.,  {Bridges} T.,  {Guenther} E.,   {Zinnecker}
  H.,  2002, \mn@doi [\aj] {10.1086/341174}, 124, 404

\bibitem[\protect\citeauthoryear{{Rochau}, {Brandner}, {Stolte}, {Gennaro},
  {Gouliermis}, {Da Rio}, {Dzyurkevich}  \& {Henning}}{{Rochau}
  et~al.}{2010}]{Roc10}
{Rochau} B.,  {Brandner} W.,  {Stolte} A.,  {Gennaro} M.,  {Gouliermis} D.,
  {Da Rio} N.,  {Dzyurkevich} N.,   {Henning} T.,  2010, \mn@doi [\apjl]
  {10.1088/2041-8205/716/1/L90}, 716, L90

\bibitem[\protect\citeauthoryear{{Scally} \& {Clarke}}{{Scally} \&
  {Clarke}}{2001}]{Sca01}
{Scally} A.,  {Clarke} C.,  2001, \mn@doi [\mnras]
  {10.1046/j.1365-8711.2001.04274.x}, 325, 449

\bibitem[\protect\citeauthoryear{{Schaefer} et~al.,}{{Schaefer}
  et~al.}{2016}]{Sch16}
{Schaefer} G.~H.,  et~al., 2016, \mn@doi [\aj] {10.3847/0004-6256/152/6/213},
  152, 213

\bibitem[\protect\citeauthoryear{{Schaller}, {Schaerer}, {Meynet}  \&
  {Maeder}}{{Schaller} et~al.}{1992}]{Sch92}
{Schaller} G.,  {Schaerer} D.,  {Meynet} G.,   {Maeder} A.,  1992, \aaps, 96,
  269

\bibitem[\protect\citeauthoryear{{Schechter}}{{Schechter}}{1976}]{Sch76}
{Schechter} P.,  1976, \mn@doi [\apj] {10.1086/154079}, 203, 297

\bibitem[\protect\citeauthoryear{{Schlafly} et~al.,}{{Schlafly}
  et~al.}{2014}]{Sch14}
{Schlafly} E.~F.,  et~al., 2014, \mn@doi [\apj] {10.1088/0004-637X/786/1/29},
  786, 29

\bibitem[\protect\citeauthoryear{{Shakura} \& {Sunyaev}}{{Shakura} \&
  {Sunyaev}}{1973}]{Sha73}
{Shakura} N.~I.,  {Sunyaev} R.~A.,  1973, \aap, 24, 337

\bibitem[\protect\citeauthoryear{{Sherry}, {Walter}, {Wolk}  \&
  {Adams}}{{Sherry} et~al.}{2008}]{She08}
{Sherry} W.~H.,  {Walter} F.~M.,  {Wolk} S.~J.,   {Adams} N.~R.,  2008, \mn@doi
  [\aj] {10.1088/0004-6256/135/4/1616}, 135, 1616

\bibitem[\protect\citeauthoryear{{Solomon}, {Rivolo}, {Barrett}  \&
  {Yahil}}{{Solomon} et~al.}{1987}]{Sol87}
{Solomon} P.~M.,  {Rivolo} A.~R.,  {Barrett} J.,   {Yahil} A.,  1987, \mn@doi
  [\apj] {10.1086/165493}, 319, 730

\bibitem[\protect\citeauthoryear{{Stolte} et~al.,}{{Stolte}
  et~al.}{2010}]{Sto10}
{Stolte} A.,  et~al., 2010, \mn@doi [\apj] {10.1088/0004-637X/718/2/810}, 718,
  810

\bibitem[\protect\citeauthoryear{{St{\"o}rzer} \& {Hollenbach}}{{St{\"o}rzer}
  \& {Hollenbach}}{1999}]{Sto99}
{St{\"o}rzer} H.,  {Hollenbach} D.,  1999, \mn@doi [\apj] {10.1086/307055},
  515, 669

\bibitem[\protect\citeauthoryear{{Tazzari} et~al.,}{{Tazzari}
  et~al.}{2017}]{Taz17}
{Tazzari} M.,  et~al., 2017, \mn@doi [\aap] {10.1051/0004-6361/201730890}, 606,
  A88

\bibitem[\protect\citeauthoryear{{Testi}, {Natta}, {Scholz}, {Tazzari}, {Ricci}
   \& {de Gregorio Monsalvo}}{{Testi} et~al.}{2016}]{Tes16}
{Testi} L.,  {Natta} A.,  {Scholz} A.,  {Tazzari} M.,  {Ricci} L.,   {de
  Gregorio Monsalvo} I.,  2016, \mn@doi [\aap] {10.1051/0004-6361/201628623},
  593, A111

\bibitem[\protect\citeauthoryear{{Tobin}, {Hartmann}, {Furesz}, {Mateo}  \&
  {Megeath}}{{Tobin} et~al.}{2009}]{Tob09}
{Tobin} J.~J.,  {Hartmann} L.,  {Furesz} G.,  {Mateo} M.,   {Megeath} S.~T.,
  2009, \mn@doi [\apj] {10.1088/0004-637X/697/2/1103}, 697, 1103

\bibitem[\protect\citeauthoryear{{Tripathi}, {Andrews}, {Birnstiel}  \&
  {Wilner}}{{Tripathi} et~al.}{2017}]{Tri17}
{Tripathi} A.,  {Andrews} S.~M.,  {Birnstiel} T.,   {Wilner} D.~J.,  2017,
  \mn@doi [\apj] {10.3847/1538-4357/aa7c62}, 845, 44

\bibitem[\protect\citeauthoryear{{Vincke} \& {Pfalzner}}{{Vincke} \&
  {Pfalzner}}{2016}]{Vin16}
{Vincke} K.,  {Pfalzner} S.,  2016, \mn@doi [\apj]
  {10.3847/0004-637X/828/1/48}, 828, 48

\bibitem[\protect\citeauthoryear{{Williams}, {de Geus}  \& {Blitz}}{{Williams}
  et~al.}{1994}]{Wil94}
{Williams} J.~P.,  {de Geus} E.~J.,   {Blitz} L.,  1994, \mn@doi [\apj]
  {10.1086/174279}, 428, 693

\bibitem[\protect\citeauthoryear{{Winter}, {Clarke}, {Rosotti}  \&
  {Booth}}{{Winter} et~al.}{2018}]{Win18}
{Winter} A.~J.,  {Clarke} C.~J.,  {Rosotti} G.,   {Booth} R.~A.,  2018, \mn@doi
  [\mnras] {10.1093/mnras/sty012}, 475, 2314

\bibitem[\protect\citeauthoryear{{Wright}, {Drake}, {Drew}  \& {Vink}}{{Wright}
  et~al.}{2010}]{Wri10}
{Wright} N.~J.,  {Drake} J.~J.,  {Drew} J.~E.,   {Vink} J.~S.,  2010, \mn@doi
  [\apj] {10.1088/0004-637X/713/2/871}, 713, 871

\bibitem[\protect\citeauthoryear{{Wright}, {Drew}  \& {Mohr-Smith}}{{Wright}
  et~al.}{2015}]{Wri15}
{Wright} N.~J.,  {Drew} J.~E.,   {Mohr-Smith} M.,  2015, \mn@doi [\mnras]
  {10.1093/mnras/stv323}, 449, 741

\bibitem[\protect\citeauthoryear{{Wright}, {Bouy}, {Drew}, {Sarro}, {Bertin},
  {Cuillandre}  \& {Barrado}}{{Wright} et~al.}{2016}]{Wri16}
{Wright} N.~J.,  {Bouy} H.,  {Drew} J.~E.,  {Sarro} L.~M.,  {Bertin} E.,
  {Cuillandre} J.-C.,   {Barrado} D.,  2016, \mn@doi [\mnras]
  {10.1093/mnras/stw1148}, 460, 2593

\bibitem[\protect\citeauthoryear{{de Gregorio-Monsalvo} et~al.,}{{de
  Gregorio-Monsalvo} et~al.}{2013}]{deG13}
{de Gregorio-Monsalvo} I.,  et~al., 2013, \mn@doi [\aap]
  {10.1051/0004-6361/201321603}, 557, A133

\makeatother
\end{thebibliography}


\appendix

\section{Cluster Modelling}
\label{sec:clust_models}

Below we review the assumptions made in the cases of specific clusters in order to model the FUV flux in the region, the results of which are shown in Section \ref{sec:local_environs}. 

\subsection{Cygnus OB2}
\label{sec:cygmodel}
Cygnus OB2 (Cyg OB2) is a young association in which the majority of stars formed $1$-$7$~Myr ago. It has a large population of $52$ O-stars, and an estimated total mass of $1$-$3\times 10^4 \, M_\odot$ where the largest stellar mass is $\sim 100 \, M_\odot$ \citep{Wri15}. The disc fraction as a function of FUV flux within the cluster has been investigated in detail by \citet{Gua16}. They find that disc survival rates are reduced in regions of higher $G_0$, with $40\%$ of stars hosting discs for $G_0 \sim 10^3 $ dropping to $\sim 20\%$ for $G_0 \gtrsim 10^4$. For the stellar density profile (Equation \ref{eq:rho_model}) \citet{Wri16} find a value of $\gamma = 5.8 \pm 0.5$ and $a=19.4\pm 1.9'$  which corresponds to $7.5$~pc at a distance of $1.33$~kpc. 

As Cyg OB2 is a well studied cluster, and we are therefore able to make some corrections to our calculations to take into account the modest sub-structure and   a slightly different IMF. The IMF in Cyg OB2 is found to be marginally shallower ($\xi \propto m^{-2.39\pm0.19}$) in the high-mass end. An increased population of massive stars will alter the $G_0$ estimates and we therefore adopt this shallower IMF for $m>1\,M_\odot$.  To estimate the density enhancement, we apply the results of \citep{Gua16} who used minimum spanning trees to simulate sub-clustering within Cyg OB2. We introduce a multiplicative factor to our number density profile such that the fraction of stars with number densities $>200$~pc$^{-3}$ agrees with the results shown in Figure 12 of that paper. This results in an enhancement in the number densities by a factor $\sim 12$. Similarly the FUV flux is enhanced by the reduced distance to neighbouring stars, and \citet{Gua16} find the $G_0\sim 10^4$ in the core. This only increases our $G_0$ estimates by a small factor. We present both the enhanced and non-enhanced cases. 

\subsection{Serpens Star Forming Region}

The recent study of \citet{Law17} found no significant differences in the PPD masses in the Serpens star forming region when compared to the low-density Taurus region which is of a similar age ($1-3$~Myr). This suggests that neither tidal truncation nor external photoevaporation has had a significant influence on the disc evolution in this region. 

At least two main sub-clusters are present in the Serpens region, Serpens A and B. \citet{Har07} find the radius of sub-cluster A(B) to be $\sim 0.25$($0.21$)~pc. They contain $44$ and $17$ stars respectively, while the the rest of the region contains an additional $174$, at an average number density of $\sim 2.5$~pc$^{-3}$, and this sample is complete down to  masses $ \sim 0.08 \, M_\odot$. \mbox{\citet{Eri15}} find the most massive star in Serpens to be $5.1$~$M_\odot$ located at R.A. 18~h 29~m 56.1~s and Dec. 01$^\circ$ 00' 21.7'' which places it close to the centre of Serpens A as projected onto the sky. 

We model Serpens A and B as two Plummer spheres (with $\gamma=4$ in Equation \ref{eq:rho_model}) with a maximum stellar mass of $5.1\, M_\odot$ placed at the centre of cluster A. The projected separations of the two sub-clusters is $\sim 3$~pc, which we use as our physical separation. The scale factors $a = 0.25$, $0.21$~pc are taken for A and B respectively. The mass, $M_\mathrm{clust}$, of each A and B is fixed so that the correct number of stars are found within $a$ from the centre when drawn from the IMF truncated above $5.1\,M_\odot$. We remove all stars outside of the radius $a$ from the centre of the two Plummer spheres. Serpens has an elongated, filamentary shape, and therefore we arrange the remaining stars isotropically over a rectangular box centred on Serpens B such that the total number of stars is $235$. We assume that the box has dimensions such that the two shortest sides have equal length of $2$~pc and the third has length $7$~pc. Because the number of stars in Serpens is relatively small, the approximate FUV flux experienced by those stars is dependent on the stochastic ICs. We therefore produce $100$ versions of this model and perform statistics on the full sample in Serpens A, B and the remaining population. Thus we produce a reasonable range of $G_0$ in the two cores. 

We find that all versions of these initial conditions produce a local FUV flux which is $ \ll 1 \, G_0$ in all regions of Serpens. As the interstellar value is unity, we adopt this as the floor in our FUV flux estimates. Hence the irradiation of discs due to member stars is considered to be insignificant. We choose the extremal number densities in all of our model generations as the range of $n_\mathrm{c}$.

\subsection{IM Lup in Lupus 2}

The Lupus clouds are a low-mass star forming complex located $\sim 140-200$~pc from the Sun. It is composed of multiple physically separated associations \citep[e.g.][]{Com08}. They are projected along the sky against the Scorpius-Centaurus OB association (Sco OB2), which is a distance of $\sim 140$~pc from the Sun and comprised of several spatially separated groups with varying ages. The stellar components of Sco OB2 are $\sim 5-16$~Myr old, with masses up to $\sim 20\, M_\odot$ and an approximate IMF with $\xi \propto m^{-2.6}$ at the high-mass end \citep{Pre02}. The number of OB members in close proximity to Lupus suggests a larger ambient field of UV radiation than other comparable low-mass star forming regions. 

\citet[][see also \citealt{Haw17b}]{Cle16} studied the gas and dust structure of the disc around IM Lup, a $1\, M_\odot$ young ($\lesssim 1 $~Myr) M0 type star associated with the Lupus 2 cloud, $\sim 160$~pc from the Sun. They make an estimate of the local $G_0 \sim 2.9 - 4.5$ depending on assumptions made about extinction, sufficient to alter the gas phase CO profile within the disc. 

Clearly the diffuse and clumpy Lupus region is not well suited to modelling using the same density profile as in other cases. We do not estimate the local stellar number density in the region around IM Lup directly, but instead argue that the most dense region in Lupus is Lupus 3, which is thought to have a stellar number density up to $\sim 500$~pc$^{-3}$ in the cores \citep{Nak00, Mer08}. This serves as an upper limit on the local number density around IM Lup. 

\subsection{NGC 1977}

\citet{Kim16} reported the discovery of seven proplyds in NGC 1977, a region which experiences much weaker FUV fields than the core of the location of the classic proplyds in the core of the ONC, with a $G_0$ value $10-30$ times lower. NGC 1977 is located at the interface between the Orion molecular cloud and the H~II region S279 \citep{Kut76}. The ionising source in this region is a B1 V star, HD 37018 (42 Ori), which is estimated to have a mass of $10$~$M_\odot$. Thus the FUV flux at the distance of the proplyds (at separations of $\sim 0.2$~pc from 42 Ori) is estimated to be $\sim 3000 G_0$ by \citet{Kim16}. 

In total the region contains $\sim 170$ young stellar objects and $3$ young B stars within a region of radius $\sim 10$', or $\sim 1$~pc \citep{Pet08}. We therefore estimate the stellar density in the region to be $\sim 40$~pc$^{-3}$.

\subsection{$\sigma$ Orionis}

The disc population of $\sigma$ Orionis ($\sigma$ Ori), a $\sim 3$~Myr old cluster at a distance of $350$-$440$~pc \citep{May08,She08}, has been surveyed using both Herschel/PACS \citep{Mau16} and ALMA \citep{Ans17}. \citet{Mau16} report that $23\%$ of the $142$ T-Tauri stars in the dense core of radius $\sim 20'$ (or $\sim 2$~pc) are disc-hosting candidates, while the disc fraction outside this core out to $\sim 30'$ is $42\%$. \citet{Ans17} also conclude that the dust mass within discs decreases with stars with closer proximity to the central massive star.

Observed stellar masses in $\sigma$ Ori range from the O9 V star $\sigma$ Ori A with mass  $\sim 17 \, M_\odot$ down to brown dwarves with a minimum mass $\sim 0.033 \, M_\odot$ \citep{Cab08a}. In fact $\sigma$ Ori A is part of a triple system, a spectroscopic binary previously considered to be a single star with components of mass $17$ and $12.8 \, M_\odot$, and a B0.5 V star $\sigma$ Ori B at a separation of $0.25''$ and mass $\sim 11.5$ \citep{Sch16}. For modelling purposes, as these components have similar masses, we place all of these stars in the centre of the cluster with a separation of $100$~au for the wide binary (period $\sim 150$~yrs) and a distance of $8$~au for the tight binary (period $\sim 150$~days). 

The density profile in $\sigma$ Ori was modelled by \citet{Cab08}, where the surface density distribution is found to be well fit by a power law $\propto r_\mathrm{c}^{-1}$ in the core, with a steeper slope of $\propto r_\mathrm{c}^{-1.3}$ between $21'$ and $30'$ from the centre of the cluster. However, we find that by allowing small values of $a$ and $\gamma$ we can also fit this profile sufficiently with our assumed density profile in Equation \ref{eq:rho_model}. Small values of $\gamma\leq 2$ are acceptable because, although we don't have a value for $r_\mathrm{eff}$, we truncate the cluster outside $30'$ ($r_\mathrm{t} \approx 3$~pc).  We fit the mass of the cluster using the average mass obtained from Equation \ref{eq:imf} between $0.08$ and $17\, M_\odot$ and the total number of members in the Mayrit catalogue, $338$ \citep{Cab08a}. Our density profile is such that the same number of sources can be found within $3$~pc. We note that a number of these candidates might be falsely associated with the cluster, and that the catalogue includes a number of brown dwarves. However for our purposes of number density and FUV flux calculations this approximation is sufficient.

 \subsection{$\lambda$ Orionis}
 
The $\lambda$ Orionis ($\lambda$ Ori) star forming region is an OB association at a distance of around $420$~pc from the Sun \citep{Sch14}. It began forming stars $\sim 5$~Myr ago, and is located inside a shell-like structure of dust and gas which is thought to be the result of a supernova explosion $\sim 1$~Myr ago \citep{Dol01, Lee15} . Its proximity makes it a good candidate for studying disc populations, and previously \citet{Her10} have used data from the \textit{Spitzer Space Telescope} to observe disc fractions of $\sim 20\%$ around M-type stars. However, at present there are no studies which establish the dependence of disc properties on location within the association. 
 
\citet{Dol01} report the masses of the $20$ OB stars associated with $\lambda$ Ori, of which the most massive HD 36861 (also known as $\lambda$ Ori, with spectral type O8 III) has a mass of $26.8 \, M_\odot$, and lies in the centre of the region. 

With regards to the spatial distribution of the stars, it is noted that it is possible that the region formed in a flattened molecular cloud, and therefore does not have 3D symmetry \citep{Mad87}. Also, the presence of the actively star forming clouds B30 and B35 at a distance $\sim 2^\circ$ from the central star means that the projected surface density is not isotropic. In order to model the region close to $\lambda$ Ori, we truncate our stellar distribution outside $15$~pc. We then fit our surface density profile, Equation \ref{eq:surf_model}, to that obtained by \citeauthor{Dol01} in that range. Applying our IMF truncated at $26.8 \, M_\odot$, we find that the total mass up to $r_\mathrm{t}=15$~pc is $214\, M_\odot$, which is approximately consistent with the $450-600 \, M_\odot$ estimate of \citeauthor{Dol01} for the whole region. As in the case of $\sigma$ Ori, we allow small values of $a$ and $\gamma$, and obtain a similar density profile.

\subsection{NGC 2024}

NGC 2024 (also known as Orion B) is an HII region in the Orion star forming complex around $415$~pc away \citep{Ant82}. It is thought to be $\sim 0.5- 1$~Myr old \citep{Lev06, Get14}, although there is considerable extinction of $\gtrsim  27$~magnitudes due to dust in the region \citep{Len04}. 

The region is known to contain $\sim 300$ sources \citep[e.g.][]{Mey96}, of which around $85\%$ show evidence of hosting a disc \citep{Hai00, Hai01}. \citet{Man15} studied this PPD population and found no evidence of disc mass dependence on the projected distance from the massive star IRS 2b, which they attribute either to the youth of the cluster or the insufficient flux from the ionising source. The spectral type of IRS 2b itself is not well constrained, with \citet{Bik03} concluding it has spectral type O8 V-B2 V ($15-25 \, M_\odot$). Similarly, the region has a clumpy dust distribution and the extent of the extinction is not well characterised. 

\citet{Lad99} estimated the radius of the region in which there are $300$ stars to be $0.9$~pc, and the average stellar surface density in the area to be $179$~pc$^{-2}$. The central $0.1$~pc (projected from the centre) encompasses $50$ stars, and therefore the association has a central surface density of $1600$~pc$^{-2}$. While this is not sufficient to fit a full density profile, we assume the latter number density represents the central value, and fit associated values of $\gamma$ and $a$.

We model two versions of NGC 2024, with $m_\mathrm{max}=15\, M_\odot$, $25 \, M_\odot$. Although IRS 2b does not lie directly in the centre of the cluster, we place it there for simplicity. Given that dynamical mass segregation can occur on short time-scales \citep[e.g.][]{All09} this is a reasonable assumption for the long-term properties of the environment.

\section{Notes on Specific Flux-Density Contours}
\label{sec:specdisc}
The results for certain clusters presented in Figure \ref{fig:nvsf} require discussion where particular simplifying assumptions have been made. These cases are discussed below.

\subsection{Wd 1}

In all of the six examples for which density profiles are taken directly from the review of \citet{Zwa10}, with the exception of the ONC and Westerlund 1 (Wd 1), the maximum stellar masses are taken to be such that $84\%$ of clusters of equal mass are expected to contain a higher mass star ( $m_\mathrm{max} = m_\mathrm{max}^{-\sigma}$). In the case of the ONC we use the observed maximum stellar mass which coincides with this value. For Wd 1 we find $m_\mathrm{max}^{-\sigma} \approx 114\, M_\odot$, which is the only case which is greater than the upper mass limit for our stellar atmosphere models. We have therefore used this upper limit, $m_\mathrm{max} = 100$. This is a further underestimate of the flux in the region. However we find that for massive clusters where the upper limit of the IMF is relatively well sampled, the FUV flux in the cluster is less sensitive to $m_\mathrm{max}$. In Figure \ref{fig:nvsf}, all the massive clusters follow contours in the parameter space within an order of magnitude of each other, particularly in the most dense regions. Hence our decision for Wd 1 is justified.

\subsection{Cygnus OB2}

For the Cygnus OB2 association, we show two contours in Figure \ref{fig:nvsf} (both in brown, enhanced density marked by triangles), for the first of which we simply use the results as implied by our density profile without any sub-structure. The second takes the same results normalised to reflect the maximum densities and fluxes obtained by \citet{Gua16}, wherein the considerable sub-structure observed in the association is accounted for. While this is a crude approximation, we find that the factors $\sim 12$ and $\sim 1.2$ for number density and flux respectively. This suggests that the effect of sub-structure enhances number density more than the local $G_0$ values. We would expect this as on large scales given that most stars will not have any significant reduction in the distance to the most massive stellar components of the cluster which make up the dominant contribution to the FUV flux.

\subsection{NGC 2024}

Finally, the association NGC 2024 is also represented by two contours. Because of the observational complications in that region, the stellar masses are not well constrained and hence we have produced two models for $m_\mathrm{max}= 15 \,M_\odot$, $25\, M_\odot$. This represents a range of likely fluxes in the region, although the ionization in the region is consistent with a source closer to $\sim 25 M_\odot$ \citep{Bik03}. Given the difficulty modelling the clumpy dust distribution, we do not account for extinction in the region, which may somewhat reduce flux estimates. However, as NGC 2024 is contained within a small region $\sim 0.9$~pc in radius, we expect the range of fluxes suggested by the two contours without extinction to be reasonable.

\section{Particle Number Convergence}
\label{sec:numconv}

\begin{figure}
	\includegraphics[width=\columnwidth]{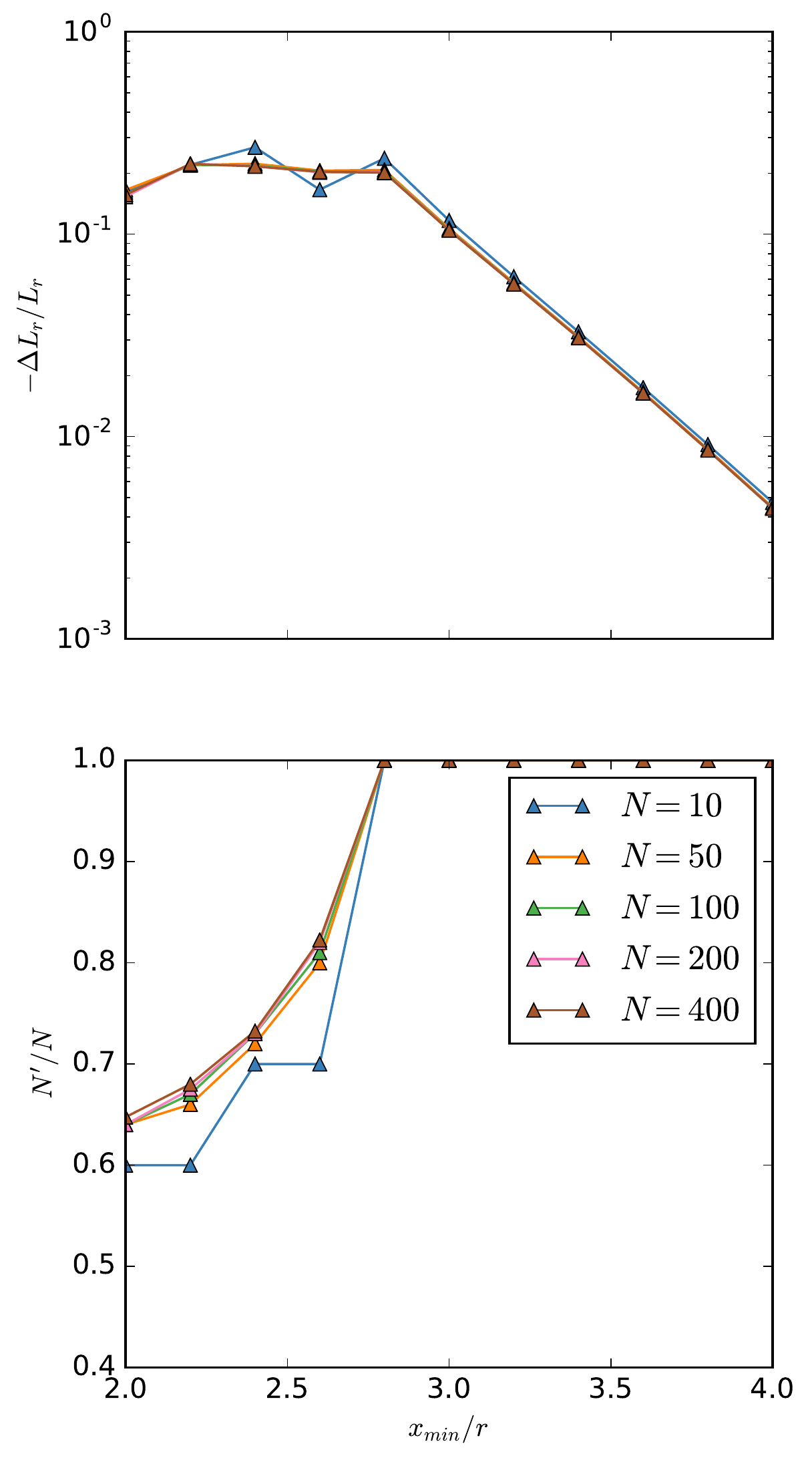}
    \caption{Results of the perturbation of an orbiting ring of test particles by an equal mass host in a coplanar, prograde, parabolic encounter. Top: mean fractional angular momentum loss of those particles that remain bound to the original host star. Bottom: fraction of particles which remain bound to the host $N'/N$. Results are shown for varying numbers of test particles, $N=10$, $50$, $100$, $200$ and $400$. }
    \label{fig:converge}
\end{figure}

In Section \ref{sec:tidal} we use test particle ring simulations to calculate the angle-averaged change of disc outer radius due to an arbitrary encounter. To confirm that the choice of the number of particles in our simulations ($N=200$) is sufficient, we run a convergence test. We calculate the change of angular momentum and the fraction of surviving particles in a ring of test particles perturbed by an equal mass companion on a parabolic, coplanar, prograde trajectory \citep[see][]{Hal96, Win18}. We do this for $N= 10$, $50$, $100$, $200$ and $400$ to confirm that our results are not resolution dependent.

We present the results of the convergence test in Figure \ref{fig:converge}, which suggests that $N=200$ is sufficient for our purposes. There is no significant change in the results until $N<50$. Further, the particle rings only contribute to the outer radius calculation if the surviving particle fraction $N'/N >0.9$ (see Section \ref{sec:rout_def}) which limits concerns about convergence to more distant encounters. In conclusion, we do not find that our results are resolution limited.

\section{Fitting Formulae}
\label{sec:fitform}

We consider the nature of the fitting formulae which we apply to the numerical results for disc truncation radii. It turns out that an appropriate general form for a model is complicated by the non-trivial dependence on  $M_2/M_1$, $e_\mathrm{pert}$ and $x_\mathrm{min}/R_\mathrm{out}$ for the contribution to the fractional change in angular momentum ${\Delta L/L}$ from various resonances. The dominant resonance in a given region of parameter space defines how the truncation radius scales locally with these variables. Creating a complete fitting function for each resonance would be both numerically challenging and of limited use for application to cluster models. Fortunately, most regions of parameter space for which encounters are expected to be important can be modelled simplistically such that the the resulting fitting formula is an accurate description of the numerical results to within $\sim 10 \%$.

Three distinct regions in $x_\mathrm{min}/R_\mathrm{out}$ space can be identified. Henceforth it is more convenient to work in reciprocal space, and we denote $R_\mathrm{out}/x_\mathrm{min} \equiv R_x$, with the associated post-encounter fractional radius $R'_\mathrm{out}/x_\mathrm{min} \equiv R'_x$. In \citet{Win18} we show that the distant encounters have an negligible influence on the disc, and we are therefore free to assume that for $R_x$ smaller than some limit, $R'_x \approx R_x$, which we call the `distant regime'. 

The `close regime' (highly penetrating encounter) is the opposite limit for which $R_x$ is large. In this regime we expect $R'_x$ to be independent of $R_x$, and therefore to be a constant for fixed $M_2/M_1$, $e_\mathrm{pert}$ (i.e. in this limit the final disc radius is independent of the disc's initial outer radius). As the angular momentum loss increases with $M_2/M_1$ and decreases with increasing $e_\mathrm{pert}$ for close encounters, we expect the opposite relationship for $R'_x$. We also find empirically that the dependence of this truncation radius on $e_\mathrm{pert}$ decreases as $M_2/M_1$ increases. 

\begin{figure}
	\includegraphics[width=\columnwidth]{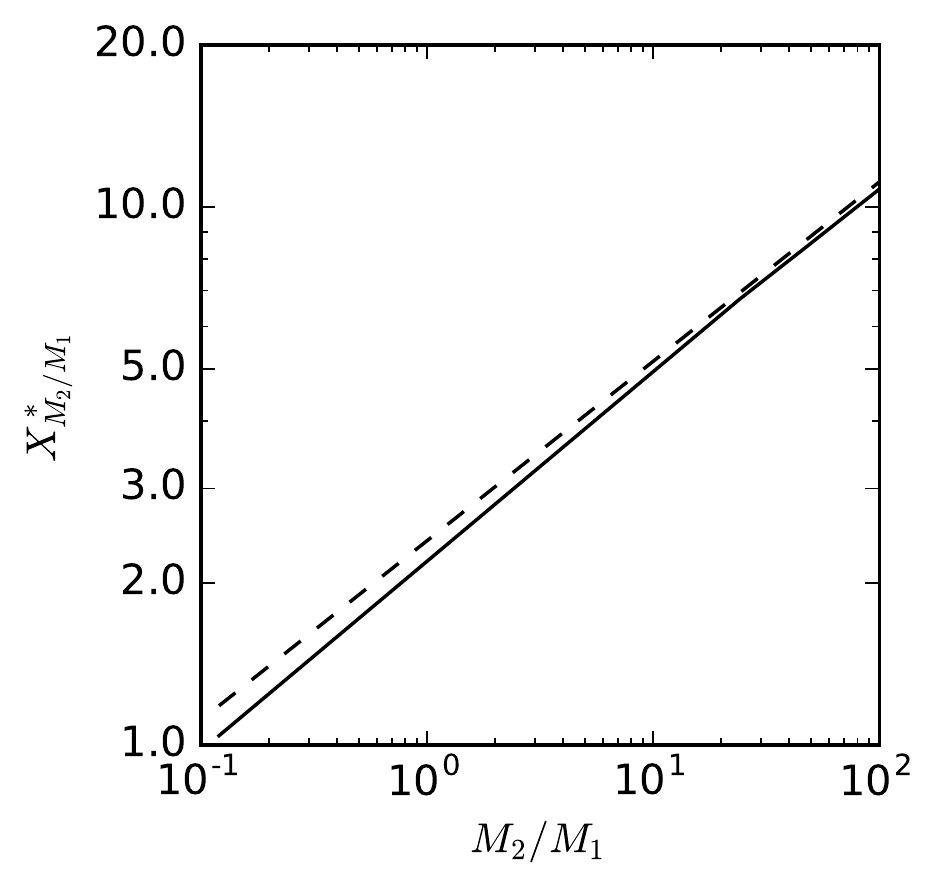}
    \caption{The lower limit of the fractional closest approach distance $1/R_x$ for which the linearised equations apply, defined to be where $\Delta L/L =0.1$ at the outer edge of the disc. The dashed line is the approximate value from Equation \ref{eq:xlim}, while the solid line is the value obtained directly from the linearised equations \citep{Win18}.}
    \label{fig:xlims}
\end{figure}

In order to continue we identify a useful quantity which we use to generalise results for unit mass ratio to arbitrary $M_2/M_1$.  In \citet{Win18} we indicate the limiting distance for a closest approach above which linearised equations are applicable: $1/R_x > X_{M_2/M_1}^*$, which is a function of the ratio of the perturbing to host masses. This is defined by the value of $R_x$ for which $\Delta L/L=0.1$ at the outer edge of the disc. It turns out that the results of the linear analysis at a range of mass ratios can be fitted by
\begin{equation}
\label{eq:xlim}
X^{*}_{M_2/M_1} \approx 2.4(M_2/M_1)^{1/3}
\end{equation} which is consistent with the findings of \citet{Vin16}. Equation \ref{eq:xlim} is plotted against the theoretical value obtained directly from the linearised equations in Figure \ref{fig:xlims} \citep{Win18} This quantity defines the ratio of $x_\mathrm{min}$ to $R_\mathrm{out}$ within which encounters are significant and therefore provides an approximate mapping between results for the $M_2/M_1=1$ case and a general perturbing mass ratio. We define
\begin{equation}
\label{eq:fdef}
f \equiv X^{*}_{1}/ X^{*}_{M_2/M_1} \approx (M_2/M_1)^{-1/3}
\end{equation} such that an encounter with closest approach $x_\mathrm{min}$ in the case that $M_2/M_1$ is not equal
to unity is deemed to be approximately dynamically equivalent to
an encounter with  closest approach $f x_\mathrm{min}$ in the case that $M_2=M_1$.


With these definitions, we define the functional form of the model in the close-regime to be
\begin{equation}
R'_{x_\mathrm{close}}\equiv \phi_1  {e_\mathrm{pert}}^{f \phi_2 } \cdot f \left( \frac {M_2}{M_1} \right)^{\phi_3}
\end{equation} where $\phi_i$ are fitting constants, $\phi_{1,2}>0$. The quantity $\phi_1$ represents the limiting value of $R_x$ for unit mass ratio and a parabolic orbit, and therefore for a parabolic orbit of arbitrary mass ratio we would expect $R'_x = f \phi_1$ according to the argument set out above. However, we have included an additional correction factor dependent on $M_2/M_1$. This is because our unmodified scale factor $f$ is based on the mass dependence of the $m=2$ ILR, which is not the dominant resonance excited in the disc for extremely close encounters. While this in some respects makes our definition of $f$ redundant, we expect this correction factor to be small ($|\phi_3| \ll 1$), and $f$ is still meaningful in relating the scaling of our composite solutions. We have additionally simplified our model by making assumptions about how the dependence on eccentricity is related to the mass ratio, scaling $\phi_2$ by $f$. 

The functional form of the `intermediate region' (between the region of negligible truncation and tidal truncation to a fixed fraction of the closest approach) is extremely complex. However, we find a much simplified linear prescription for the new outer radius to be acceptable:

\begin{equation}
R'_{x_\mathrm{inter.}} \equiv  (1-\psi_1  e_\mathrm{pert}^{-\psi_2}) R_x  +f \psi_1 \psi_3  e_\mathrm{pert}^{-\psi_2}
\end{equation} where $\psi_i>0$ are fitting constants, and $\psi_1 <1$. 

Our full model for the post-encounter radius is then
\begin{equation}
\label{eq:fullmod}
R'_x = \mathrm{min} \left\{ R_x, R'_{x_\mathrm{inter.}}, R'_{x_\mathrm{close}} \right\}
\end{equation} fully defined by the six fitting parameters $\phi_{i=1,2,3}, \psi_{i=1,2,3}$. We apply the Python implementation for MCMC, \textsc{emcee} \citep{For13} to fit our model and establish errors in the $M_2/M_1=1$ case for five of these parameters. However for $\phi_3$ we simply refit for a high mass ratio example $M_2/M_1=10$, using the rest of the parameters as found from the $M_2/M_1=1$ case.

\section{Encounter Rate Parameterisation}
\label{sec:encrate}

In the discussion of the significance of encounters on a disc, the usual approach is to ask some variation on the question `what is the probability that a disc experiences an encounter closer than some separation $x_\mathrm{min}$?' \citep[e.g.][]{Bin87, Ost94, Duk12, Mun15}.  However, this question crucially depends on the effective number density of the stellar population $n_\mathrm{eff}$ which is likely to be dependent both on sub-structure evolution and spatial location within the cluster \citep{Cra13}. Instead of trying to model the global evolution of a stellar population with a spatially dependent number density distribution, we consider local conditions for simplicity. On the issue of sub-structure \citet{Cra13} found that, even for a modest fractal dimension $D$, the \textit{overall} number of close encounters during the lifetime of the cluster can become enhanced by a factor of a few, even though the sub-structure is eliminated over a crossing time. Therefore this should be considered if we want an accurate estimate of the degree to which stellar encounters are important.

The differential encounter rate is normally expressed in terms of the impact parameter $b$, which can be related to the closest approach and relative speed in the limit of distant separation $v_\infty$:
$$
b^2 = x^2_\mathrm{min} \left( 1 + \frac{2GM_\mathrm{tot}}{v^2_\infty x_\mathrm{min}}\right)
$$ For simplicity, we consider a cluster with a Maxwellian velocity distribution, dispersion $\sigma_v^2$. We define $V^2\equiv v_\infty^2/4\sigma_v^2$, then following \citet{Ost94} the differential encounter rate for a cluster with normalised IMF $\xi$ is
\begin{equation}
\label{eq:encrate}
\mathrm{d}\Gamma = \gamma(x_\mathrm{min}, V, M_2) \, \mathrm{d}x_\mathrm{min} \, \mathrm{d}V^2 \, \mathrm{d}M_2 
\end{equation} where we define $\gamma$:
$$
\gamma \equiv \frac{2\sqrt{\pi} G M_\mathrm{tot} n_\mathrm{eff}}{\sigma_v} \left(  1+\frac{4\sigma_v^2 x_\mathrm{min}V^2}{GM_\mathrm{tot}} \right)\exp(-V^2) \xi (M_2 )
$$ and $M_\mathrm{tot}$ is the combined mass of the host star $M_1$ and that of the perturber, $M_2$. In the case that we are considering a cluster comprised stars of a single mass $\bar{m}$, then the IMF becomes a delta-function $\xi(m)=\delta(m-\bar{m})$.

From \citet{Cra13} the effective number density is linked to the fractal dimension by 
$$
n_\mathrm{eff} = n_{\mathrm{c}}\cdot 2^{(3-D)(g-1)}
$$ where $g$ is the number of fractal generations and $n_\mathrm{c}$ is the number density where there is no sub-structure ($D=3$). There is a degree of arbitrariness to the number of fractal generations, but it is estimated to be 
$$
g = \frac{\ln(2N_\mathrm{c})}{\ln(8)} +1 +s_2(D)
$$ where $N_\mathrm{c}$ is the number of stars in the cluster and $s_2(D)$ is only non-zero for $D<2$, in which case it is $1$. The value of $D$ is a function of time with an uncertain evolution, however a reasonable estimate for its value in a cluster is
$$
D(t) = 3+(D_0-3)e^{-t/\tau_\mathrm{cross}}
$$ where $\tau_\mathrm{cross}$ is the crossing time of the cluster. 

Apart from $D_0$, $n_\mathrm{c}$, $\sigma_v$ and $\xi$, one further parameter needs to be assumed to link $\tau_\mathrm{cross}$, $n_\mathrm{c}$ and $g$. We choose to fix the total number of stars in the cluster $N_\mathrm{c}$. Given this, the crossing time is 
$$
\tau_\mathrm{cross} = \frac{2}{\sigma_v} \left( \frac{4\pi N_\mathrm{c}}{3n_\mathrm{c}}\right)^{1/3}
$$ Hence we have a simple time dependent model of the encounter rate at any given time given by five parameters.

\bsp	
\label{lastpage}
\end{document}